\documentclass[aps,showpacs,prl,reprint,longbibliography,superscriptaddress]{revtex4-1}

\usepackage[colorlinks=true,linkcolor=blue,citecolor=blue,urlcolor=blue]{hyperref}

\usepackage{setspace} 
\usepackage{graphicx}
\usepackage{amsmath}
\usepackage{color}
\usepackage{amsmath}
\usepackage{amssymb}
\usepackage{verbatim}
\usepackage{latexsym}
\usepackage{enumerate} 
\usepackage{bm} 

\usepackage[caption=false]{subfig}
\usepackage{multirow}
\usepackage{booktabs}

\begin{document}

\title{Topological semimetallic phase in PbO$_2$ promoted by temperature\\
}

\author{Bo Peng}
\affiliation{Cavendish Laboratory, University of Cambridge, J.\,J.\,Thomson Avenue, Cambridge CB3 0HE, United Kingdom}
\affiliation{Key Laboratory of Micro and Nano Photonic Structures (MOE), Department of Optical Science and Engineering, Fudan University, Shanghai 200433, China}
\author{Ivona Bravi\'{c}}
\affiliation{Cavendish Laboratory, University of Cambridge, J.\,J.\,Thomson Avenue, Cambridge CB3 0HE, United Kingdom}
\author{Judith L. MacManus-Driscoll} 
\affiliation{Department of Materials Science and Metallurgy, University of Cambridge, 27 Charles Babbage Road, Cambridge CB3 0FS, United Kingdom}
\author{Bartomeu Monserrat} \email{bm418@cam.ac.uk}
\affiliation{Cavendish Laboratory, University of Cambridge, J.\,J.\,Thomson Avenue, Cambridge CB3 0HE, United Kingdom}

\date{\today}

\begin{abstract}
Materials exhibiting topological order host exotic phenomena that could form the basis for novel developments in areas ranging from low-power electronics to quantum computers. The past decade has witnessed multiple experimental realizations and thousands of predictions of topological materials. 
However, it has been determined that increasing temperature destroys topological order, restricting many topological materials to very low temperatures and thus hampering practical applications. Here, we propose a material realization of temperature promoted topological order.
We show that a semiconducting oxide that has been widely used in lead-acid batteries, $\beta$-PbO$_2$, 
hosts a topological semimetallic phase driven by both thermal expansion and electron-phonon coupling upon increasing temperature. 
We identify the interplay between the quasi-two-dimensional nature of the charge distribution of the valence band with the three-dimensional nature of the charge distribution of the conduction band as the microscopic mechanism driving this unconventional temperature dependence. Thus, we propose a general principle to search for and design novel topological materials whose topological order is stabilized by increasing temperature. This provides a clear roadmap for taking topological materials from the laboratory to technological devices.
\end{abstract}

\maketitle

Materials exhibiting topological order have the potential to revolutionize modern technology. For example, Chern insulators exhibit the quantum anomalous Hall effect supporting dissipationless currents that could form the basis for low-power electronics \cite{Haldane1988,Chang2013}, and topological superconductors sustain Majorana fermions that could provide a platform for robust topological quantum computers \cite{Fu2008,Qi2009,Nadj-Perge2014,Lian2018}. The past decade has witnessed the establishement of a solid theoretical foundation of topological order, and the prediction and observation of topological phenomena in a range of materials \cite{Hasan2010,Qi2011}. 

However, topological phenomena are currently mostly restricted to very low temperatures precluding most applications. Current experiments invariably show that increasing temperature suppresses topological order \cite{Dziawa2012,Wojek2015,Xu2018,Kadykov2018,Krizman2018} and this can be rationalized with a combination of thermal expansion and electron-phonon coupling. Although electron-phonon coupling can both promote or suppress topological order \cite{Garate2013,Saha2014,Kim2015c,Monserrat2016,Antonius2016,Wang2017g,bitei_rashba_temp,moller_typeII_elph}, thermal expansion pushes materials towards the atomic limit which always exhibits a trivial band ordering, and as a result there are no examples of materials in which temperature promotes topological order. This prompts the question, is it possible to identify materials in which temperature promotes rather than suppresses topological order?

\begin{figure*}
\centering
\includegraphics[width=\linewidth]{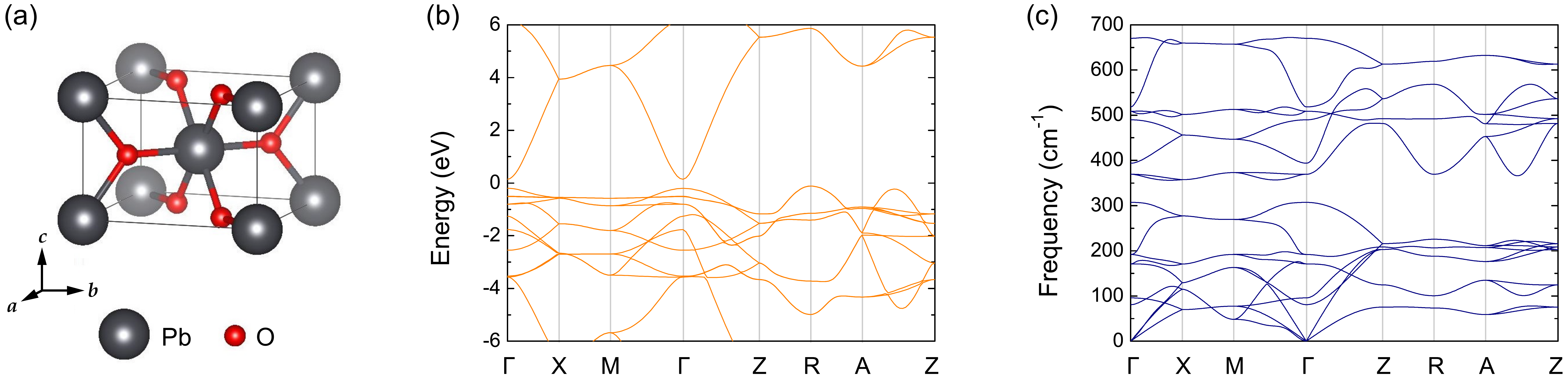}
\caption{(a) Unit cell of the crystal structure of $\beta$-PbO$_2$ with experimental lattice parameters $\textbf{\textit{a}}=4.961$\,\AA\@ and $\textbf{\textit{c}}=3.385$\,\AA\@ \cite{pbo2_exp_lattice}. (b) Electronic band structure of $\beta$-PbO$_2$ calculated using the HSE06 functional. (c) Phonon dispersion of $\beta$-PbO$_2$ calculated using the HSE06 functional.}
\label{f1} 
\end{figure*}

We present the first example of a temperature-promoted topological phase in $\beta$-PbO$_2$. With increasing temperature, \textit{both} thermal expansion and electron-phonon coupling contribute towards stabilizing a topologically ordered phase in an otherwise trivial semiconductor. 
Contrary to all known topological materials so far, we show that the coexistence of the 
quasi-two-dimensional real space charge distribution associated with the valence band of $\beta$-PbO$_2$ together with the three-dimensional charge distribution of the conduction band can promote a topological phase even under thermal expansion. 
Furthermore, the underlying microscopic mechanism we present in this work can be employed as a guiding principle to identify other materials in which the topological order is robust at high temperatures, thus opening the door for practical room temperature applications of topological matter. 

Our study is based on first-principles density functional theory (DFT) calculations performed using the Vienna \textit{ab-initio} simulation package ({\sc vasp}) \cite{Kresse1996}. The projector-augmented-wave potential is used with Pb $5d^{10}6s^26p^2$ and O $2s^22p^4$ valence states. Hybrid functional methods based on the Heyd-Scuseria-Ernzerhof (HSE06) approximation are adopted \cite{HSE1,HSE2,HSE3} with $25$\% of the exact screened Hartree-Fock exchange incorporated into the semilocal exchange within the generalized gradient approximation (GGA) in the Perdew-Burke-Ernzerhof (PBE) parametrization \cite{Perdew1996}. Based on convergence tests, we use a kinetic energy cutoff set to $500$\,eV and a $\Gamma$-centered $4\times4\times6$ $\mathbf{k}$-mesh to sample the electronic Brillouin zone. The convergence parameters for structural relaxations include an energy difference within $10^{-6}$ eV and a Hellman-Feynman force within $10^{-4}$ eV/\AA. For the calculation of surface states, we generate Wannier functions for the $s$ orbitals of lead and the $p$ orbitals of oxygen using the {\sc wannier90} package \cite{Mostofi2014} combined with surface Green's functions as implemented in the {\sc WannierTools} package \cite{Zhang2010a,Wu2018}. The phonon dispersion, thermal expansion within the quasiharmonic approximation, and the electron-phonon interaction are calculated using the finite displacement method within the nondiagonal supercell approach \cite{Lloyd-Williams2015,Monserrat2018}, using coarse $\mathbf{q}$-point grids of size $4\times4\times4$ to sample the vibrational Brillouin zone. The temperature dependence of the band structures is calculated from vibrational averages along thermal lines \cite{Monserrat2016b,Monserrat2016a}.

Lead dioxide has been one of the most widely used functional oxides since the invention of the lead-acid battery in 1860 \cite{Carr1972}. Figure~\ref{f1}(a) illustrates the most stable tetragonal $\beta$-PbO$_2$ phase (space group: $P4_2/mnm$) in which each Pb atom is at the center of an O octahedral cage. We emphasize that our subsequent analysis of $\beta$-PbO$_2$ is based on first principles DFT calculations using the HSE06 screened Coulomb hybrid density functional \cite{HSE1,HSE2,HSE3}. Compared to semilocal DFT, typically used for the study of $\beta$-PbO$_2$ \cite{Wang2017f,Chen2018a}, the HSE06 results show significantly better agreement with experimental measurements \cite{Burgio2001,Scanlon2011}, providing a correct description of the structural, electronic and vibrational properties of this material (a detailed validation study of the level of theory required to study $\beta$-PbO$_2$ can be found in the Supplemental Material \cite{Supplemental_Material_PbO2}).

As shown in Fig.\,\ref{f1}(b), PbO$_2$ is a semiconductor with the valence band maximum (VBM) at the R point and the conduction band minimum (CBM) at the $\Gamma$ point. The calculated indirect band gap of 0.24 eV and direct band gap of 0.34 eV at the $\Gamma$ point agree well with previous calculations \cite{Scanlon2011}. These results also agree well with hard X-ray photoelectron spectroscopy measurements, which indicate that $\beta$-PbO$_2$ is an instrinsic semiconductor, with the metallic behaviour exhibited by some samples arising from a partially filled conduction band due to oxygen deficiency \cite{Carr1972,Payne2009}. The phonon dispersion in Fig.\,\ref{f1}(c) shows no imaginary modes, thus indicating that the experimentally observed structure is stable at all temperatures. This is at odds with earlier DFT calculations using semilocal exchange-correlation functionals \cite{Chen2018a}, but consistent with the fact that $\beta$-PbO$_2$ is the most stable phase under normal laboratory conditions \cite{Carr1972}.  Furthermore, the HSE06 Raman active modes are in better agreement with the experimental Raman spectrum \cite{Burgio2001} than the modes calculated using semilocal DFT (see Supplemental Material \cite{Supplemental_Material_PbO2}).

\begin{figure*}
\centering
\includegraphics[width=\linewidth]{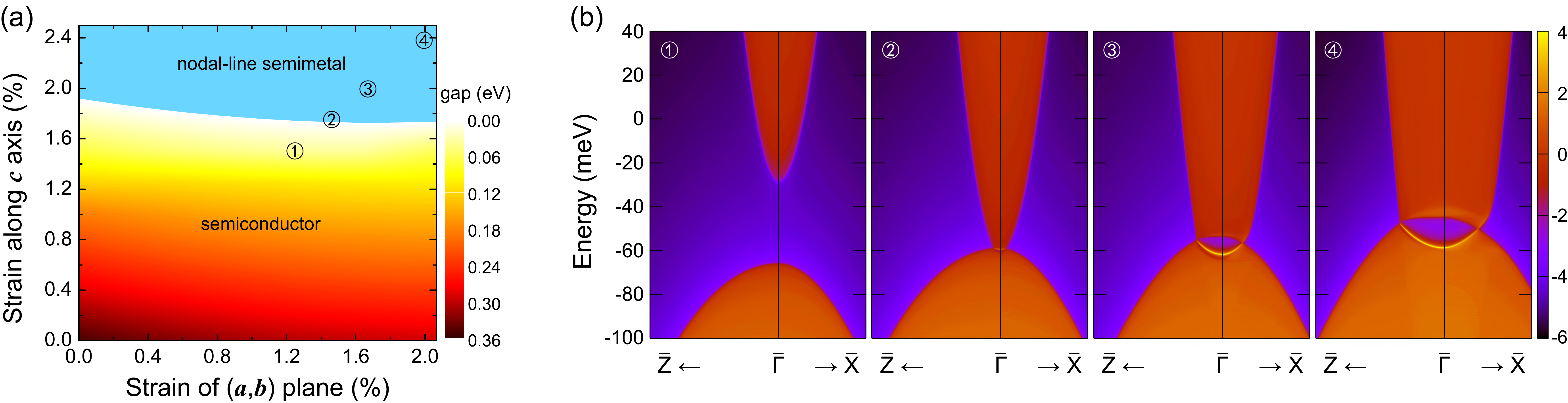}
\caption{(a) Calculated band gap at the $\Gamma$ point of $\beta$-PbO$_2$ as a function of tensile strain. (b) Effect of tensile strain on the (010) surface states of $\beta$-PbO$_2$. A warmer colour represents a higher surface contribution.}
\label{f2} 
\end{figure*}

The electronic structure of $\beta$-PbO$_2$ can be tuned from semiconducting to metallic by small variations of the lattice constants \cite{Scanlon2011,Ma2016a}. 
Interestingly, the normal band ordering at the $\Gamma$ point is inverted under tensile lattice strain, resulting in non-trivial band topology \cite{Ma2016a}. To illustrate this, we carry out band structure calculations under different tensile strains. Figure~\ref{f2}(a) presents the band gap at the $\Gamma$ point as a function of strain in the (\textbf{\textit{a}},\textbf{\textit{b}}) plane and strain along the \textbf{\textit{c}} axis. The strain conditions for which the system has normal band ordering correspond to trivial insulators with the magnitude of the band gap indicated by the red-yellow-white colour scale, whereas those with inverted band ordering are marked by the cyan colouring. Increasing the \textbf{\textit{c}} lattice constant closes the band gap, leading to a nodal-line semimetal phase (as we discuss below) at strains larger than around 1.9\%. On the other hand, the band gap is insensitive to the strain in the (\textbf{\textit{a}},\textbf{\textit{b}}) plane. Thus the \textbf{\textit{c}} axis strain is a suitable tuning parameter to control the band topology of $\beta$-PbO$_2$.

To illustrate the strain effect and estimate its magnitude, we calculate the surface states under tensile strain in Fig.\,\ref{f2}(b) on the basis of maximally localized Wannier functions \cite{Mostofi2014} and surface Green's functions \cite{Zhang2010a,Wu2018}. With an in-plane strain of 1.24\% and an out-of-plane strain of 1.50\%, the gap in the surface region decreases to 40 meV, and the conduction band moves below the Fermi energy due to the moving up of the VBM at the R point [panel 1 in Fig.\,\ref{f2}(b)]. Under a modest strain of 1.45\% and 1.75\% in the (\textbf{\textit{a}},\textbf{\textit{b}}) plane and along the \textbf{\textit{c}} axis, a topological phase transition occurs as the conduction band moves slightly below the valence band at the $\Gamma$ point [panel 2 in Fig.\,\ref{f2}(b)]. Further increasing strain leads to the formation of distinct surface states around the $\Gamma$ point [panels 3 and 4 in Fig.\,\ref{f2}(b)]. A hallmark of nodal-line semimetals is the presence of two-dimensional drumhead-like surface states connecting the projected Dirac nodal points (the projection of the bulk nodal line onto the surface plane) \cite{Burkov2011,Weng2015b}. These surface states with a divergent density of states can be detected by angle-resolved photoemission measurements \cite{Bian2016}, and may induce strong correlation phenomena such as high temperature superconductivity \cite{Kopnin2011}.

\begin{figure*}
\centering
\includegraphics[width=\linewidth]{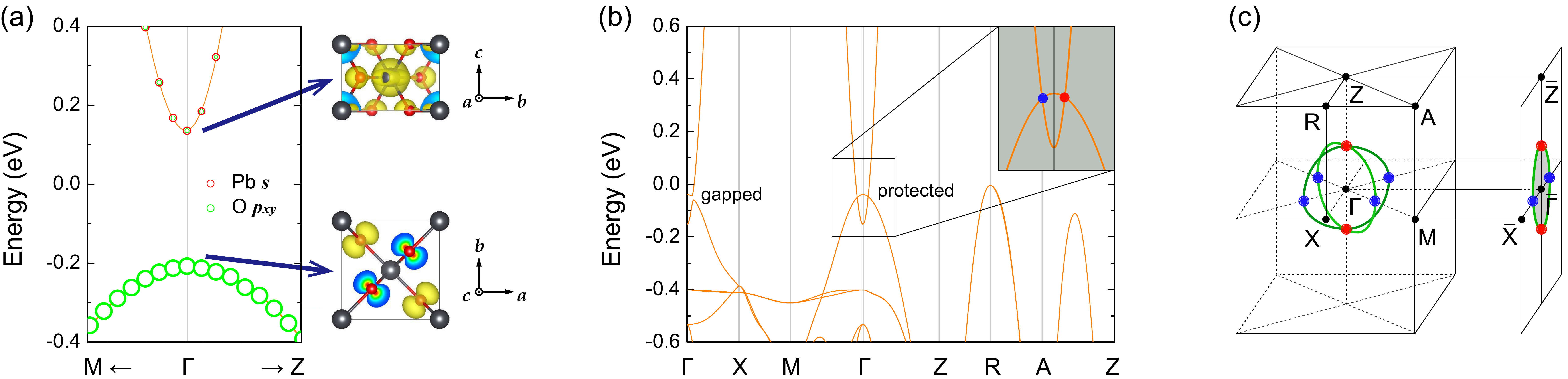}
\caption{Electronic structures for (a) unstrained and (b) strained $\beta$-PbO$_2$. Charge densities of the CBM and VBM states (isosurface = 0.008) are present in (a). (c) The bulk Brillouin zone and the surface Brillouin zone projected on the (010) surface are shown with two ring-shaped nodal lines on the (1$\bar{1}$0) and (110) planes connecting the nodal points on the M-$\Gamma$ (blue dots) and $\Gamma$-Z (red dots) high-symmetry lines.}
\label{f3} 
\end{figure*}

To rationalize the behaviour of the valence and conduction bands under tensile strain, we calculate the orbital characters near the Fermi energy. As shown in Fig.\,\ref{f3}(a), the unstrained lattice has a normal band ordering at the $\Gamma$ point with the VBM derived from the 2$p_{xy}$ orbitals of oxygen, and the CBM predominantly consists of the 6$s$ orbitals of lead that hybridize with the 2$p_{xy}$ orbitals of oxygen. We also depict the real-space charge densities corresponding to the respective VBM and CBM at the $\Gamma$ point in Fig.\,\ref{f3}(a). The VBM at the $\Gamma$ point is constituted of the O 2$p_{xy}$ atomic orbitals which are oriented along the (\textbf{\textit{a}},\textbf{\textit{b}}) plane of the primitive cell. This feature of the valence band can be seen as the establishment of a quasi-two-dimensional subsystem in the bulk. Instead, the CBM at the $\Gamma$ point resembles the diffuse Pb 6$s$ orbital which has a three-dimensional character. This means that under uniaxial strain along the $c$ direction the CBM experiences a reduced electron-electron repulsion arising from electrons of neighbouring atoms (in the \textbf{\textit{c}} direction) and as a consequence its energy decreases. In contrast, the aforementioned quasi-two-dimensional character of the valence band remains invariant under out-of-plane expansion. This leads to a relative shift of the VBM and the CBM that drives the observed band inversion. This explanation is consistent with the fact that under in-plane strain, no band inversion can be observed, which underpins the relevance of the out-of-plane anisotropy of the valence band. In this case, both valence and conduction band states experience similar stabilization effects upon expansion and therefore no crossing occurs. With larger tensile strains $\beta$-PbO$_2$ will revert to the standard behaviour in which lattice expansion leads to the topologically trivial atomic limit.

To explain the formation of nodal lines, we examine the symmetry properties of $\beta$-PbO$_2$. For a fully relaxed unstrained crystal structure, $\beta$-PbO$_2$ is a trivial indirect band gap semiconductor with the minimum direct bandgap at the $\Gamma$ point. Under tensile strain the band structure undergoes a band inversion at the $\Gamma$ point. As shown in Fig.\,\ref{f3}(b), under a modest strain of 2.07\% in the (\textbf{\textit{a}},\textbf{\textit{b}}) plane and 2.50\% along the \textbf{\textit{c}} axis [corresponding to 4 in Fig.\,\ref{f2}(a)], the band gap at the $\Gamma$ point is inverted with a band inversion energy of $0.11$\,eV, and two band crossing points are observed along the M-$\Gamma$ and $\Gamma$-Z high-symmetry lines. The crossing points belong to two ring-shaped nodal lines on the (1$\bar{1}$0) and (110) mirror symmetry planes, as shown in Fig.\,\ref{f3}(c). In the absence of spin-orbit coupling, for each mirror plane the inverted band states have mirror eigenvalues $-1$ and $+1$ respectively, thus the two crossing bands cannot hybridize with each other and are therefore not gapped, forming a nodal line on the plane. The two nodal lines on the (1$\bar{1}$0) and (110) planes are identical, which is similar to transition-metal rutile oxide PtO$_2$ \cite{Kim2019}. For comparison, the band crossing on the $\Gamma$-X high-symmetry line is not protected by mirror symmetry and therefore gapped.

We next evaluate the role of spin-orbit coupling in $\beta$-PbO$_2$. The crossing bands are comprised of the $6s$ states of the lead atoms and the $2p_{xy}$ states of the light oxygen atoms. As a consequence, none of these exhibit strong spin-orbit coupling effects, and the band structure and surface states with the inclusion of spin-orbit coupling are similar to those without spin-orbit coupling. Technically, spin-orbit coupling breaks the SU(2) symmetry, generally gapping the nodal line with only one pair of Dirac points along the $\Gamma$-Z line remaining, as has been previously observed \cite{Wang2017f}. The Dirac points are protected by the fourfold screw rotation $\tilde{C}_{4z}$ symmetry, which is a fourfold rotation about the \textit{z} axis, followed by a translation by (\textbf{\textit{a}}/2,\textbf{\textit{a}}/2,\textbf{\textit{c}}/2) \cite{Kim2019}. Thus the nodal-line semimetal evolves into a Dirac semimetal with the inclusion of spin-orbit coupling. Nonetheless, the weak spin-orbit coupling leads to a small gap of only $1.05$\,meV and the resulting physics resembles that of a nodal-line semimetal for temperatures over $12.2$\,K in which the bands merge due to the temperature-induced broadening (for further details on the spin-orbit coupling results, see the Supplemental Material \cite{Supplemental_Material_PbO2}). 
The same physics holds for $GW_0$ band structure calculations after taking exchange and correlation interactions of quasiparticles into consideration (for details on the $GW_0$ results, see the Supplemental Material \cite{Supplemental_Material_PbO2}).

\begin{figure}
\centering
\includegraphics[width=\linewidth]{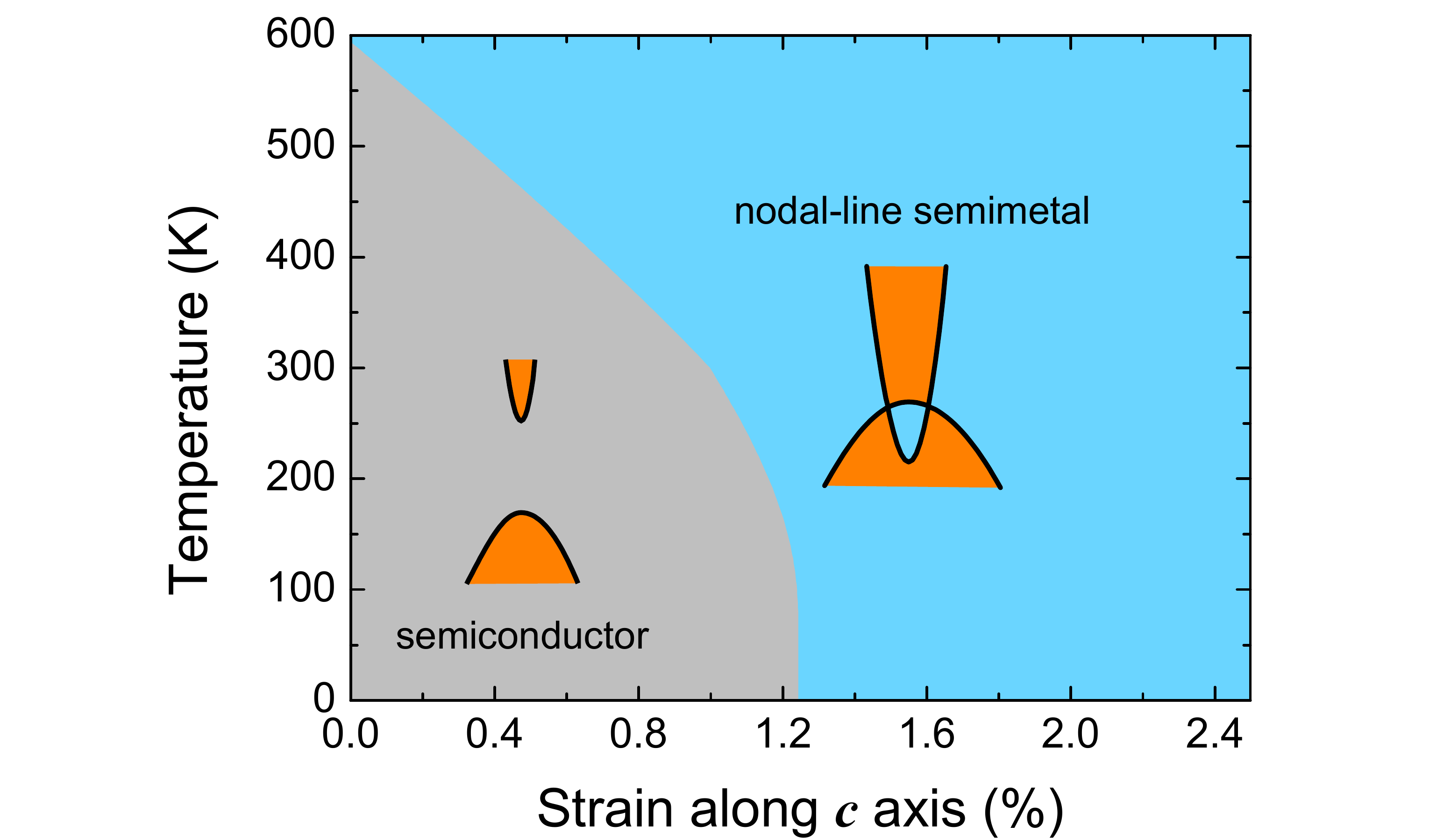}
\caption{Calculated strain-temperature phase diagram of $\beta$-PbO$_2$.}
\label{f4} 
\end{figure}

The behaviour of $\beta$-PbO$_2$ under tensile strain discussed above indicates that, contrary to all known topological materials, thermal expansion should promote topological order in $\beta$-PbO$_2$.
In addition to thermal expansion, increasing temperature also enhances electron-phonon coupling. We find that electron-phonon coupling has a contribution similar to that of thermal expansion to the temperature induced band inversion of $\beta$-PbO$_2$. Figure~\ref{f4} demonstrates the evolution of band topology as a function of temperature including both thermal expansion and electron-phonon coupling contributions, which promote the topological phase at high temperature. For unstrained PbO$_2$, the band gap closes above a temperature of $595$\,K, which is close to the temperature at which $\beta$-PbO$_2$ becomes unstable \cite{Carr1972}. Above a strain of $1.26$\%, the system can be driven into a topological phase even at $0$\,K, and in the intermediate strain regime the temperature promoted topological phase can be easily tuned to any intermediate temperature. 
We note that at $0$\,K the topological phase appears at a $\textbf{\textit{c}}$ strain of $1.26$\%, significantly lower than the required strain of around $1.9$\% at the static lattice level [see Fig.\,\ref{f2}(a)]. This difference is a purely quantum effect arising from the zero-point motion of the lattice.

The unusual temperature promoted topological order predicted in $\beta$-PbO$_2$ could be observed experimentally by exploiting the sensitivity of the transition temperature to the $\textbf{\textit{c}}$ axis strain, which would enable the tuning of the transition temperature anywhere in the range $0$--$595$\,K. Both uniaxial tensile strain along the $\textbf{\textit{c}}$ axis, or three-dimensional tensile strain using a stoa-like structure consisting of stylobate (substrate) and column (matrix) with appropriately matched lattice constants, could be used to scan over the phase diagram of $\beta$-PbO$_2$. 

The relatively low melting temperature of $\beta$-PbO$_2$ \cite{Carr1972} means that it is possible to grow highly crystalline films. The appropriate substrate and matrix materials can be chosen from candidates with lattice constants in the appropriate region of the diagram in Fig.\,\ref{f2}(a). Using rutile TiO$_2$ as a substrate, with an in-plane lattice constant of $\textbf{\textit{a}}=4.594$\,\AA, would provide in-plane compression of $\beta$-PbO$_2$ with an associated out-of-plane tension. Epitaxial growth of structurally analogous SnO$_2$ has been demonstrated on multiple orientations of a columbite CoNb$_2$O$_6$ substrate \cite{sno2_epitaxial_growth}. These substrates could be good starting points for exploring the topological phase diagram of $\beta$-PbO$_2$.

We also note that $\beta$-PbO$_2$ exhibits intrinsic defects that place the Fermi energy within the conduction bands \cite{Carr1972,Payne2009}. This implies that the energy of the surface states will not coincide with the Fermi energy, and therefore the topological surface states will coexist with other non-topological surface states. Nonetheless, we remark that this is a common feature of many topological materials (see, e.g., Refs.\,\cite{Chen2009_bi2te3,Lou2018_zrb2}) that, although detrimental for the observation of isolated topological surface states, it facilitates the use of techniques such as angle-resolved photoemission spectroscopy for the detection of such states.

In summary, we have predicted that with a combination of thermal expansion and electron-phonon couping, $\beta$-PbO$_2$ transforms from a trivial semiconductor at low temperatures to a topological semimetal at high temperatures. This behaviour contrasts with that of all known topological materials, most of which cannot hold topological order at high temperatures, with notorious examples including Chern insulators and topological superconductors which are currently restricted to few Kelvins at most. Our work proposes a microscopic picture of the chemical and physical mechanisms behind the interplay between topological order and temperature, and therefore provides a solid platform to search for novel topological materials whose topological order is robust at high temperatures. Such understanding paves the way for taking topological materials from the laboratory to technological devices.

\acknowledgements

I.B. and B.M. acknowledge support from the Winton Programme for the Physics of Sustainability. B.M. also acknowledges support from Robinson College, Cambridge, and the Cambridge Philosophical Society for a Henslow Research Fellowship. Some of the calculations were performed using resources provided by the Cambridge Tier-2 system operated by the University of Cambridge Research Computing Service (http://www.hpc.cam.ac.uk) funded by EPSRC Tier-2 capital grant EP/P020259/1.

\bibliography{new}

\begin{thebibliography}{61}%
\makeatletter
\providecommand \@ifxundefined [1]{%
 \@ifx{#1\undefined}
}%
\providecommand \@ifnum [1]{%
 \ifnum #1\expandafter \@firstoftwo
 \else \expandafter \@secondoftwo
 \fi
}%
\providecommand \@ifx [1]{%
 \ifx #1\expandafter \@firstoftwo
 \else \expandafter \@secondoftwo
 \fi
}%
\providecommand \natexlab [1]{#1}%
\providecommand \enquote  [1]{``#1''}%
\providecommand \bibnamefont  [1]{#1}%
\providecommand \bibfnamefont [1]{#1}%
\providecommand \citenamefont [1]{#1}%
\providecommand \href@noop [0]{\@secondoftwo}%
\providecommand \href [0]{\begingroup \@sanitize@url \@href}%
\providecommand \@href[1]{\@@startlink{#1}\@@href}%
\providecommand \@@href[1]{\endgroup#1\@@endlink}%
\providecommand \@sanitize@url [0]{\catcode `\\12\catcode `\$12\catcode
  `\&12\catcode `\#12\catcode `\^12\catcode `\_12\catcode `\%12\relax}%
\providecommand \@@startlink[1]{}%
\providecommand \@@endlink[0]{}%
\providecommand \url  [0]{\begingroup\@sanitize@url \@url }%
\providecommand \@url [1]{\endgroup\@href {#1}{\urlprefix }}%
\providecommand \urlprefix  [0]{URL }%
\providecommand \Eprint [0]{\href }%
\providecommand \doibase [0]{http://dx.doi.org/}%
\providecommand \selectlanguage [0]{\@gobble}%
\providecommand \bibinfo  [0]{\@secondoftwo}%
\providecommand \bibfield  [0]{\@secondoftwo}%
\providecommand \translation [1]{[#1]}%
\providecommand \BibitemOpen [0]{}%
\providecommand \bibitemStop [0]{}%
\providecommand \bibitemNoStop [0]{.\EOS\space}%
\providecommand \EOS [0]{\spacefactor3000\relax}%
\providecommand \BibitemShut  [1]{\csname bibitem#1\endcsname}%
\let\auto@bib@innerbib\@empty
\bibitem [{\citenamefont {Haldane}(1988)}]{Haldane1988}%
  \BibitemOpen
  \bibfield  {author} {\bibinfo {author} {\bibfnamefont {F.~D.~M.}\
  \bibnamefont {Haldane}},\ }\bibfield  {title} {\enquote {\bibinfo {title}
  {{{Model for a Quantum Hall Effect without Landau Levels: Condensed-Matter
  Realization of the ``Parity Anomaly"}}},}\ }\href {\doibase
  10.1103/PhysRevLett.61.2015} {\bibfield  {journal} {\bibinfo  {journal}
  {Phys. Rev. Lett.}\ }\textbf {\bibinfo {volume} {61}},\ \bibinfo {pages}
  {2015--2018} (\bibinfo {year} {1988})}\BibitemShut {NoStop}%
\bibitem [{\citenamefont {Chang}\ \emph {et~al.}(2013)\citenamefont {Chang},
  \citenamefont {Zhang}, \citenamefont {Feng}, \citenamefont {Shen},
  \citenamefont {Zhang}, \citenamefont {Guo}, \citenamefont {Li}, \citenamefont
  {Ou}, \citenamefont {Wei}, \citenamefont {Wang}, \citenamefont {Ji},
  \citenamefont {Feng}, \citenamefont {Ji}, \citenamefont {Chen}, \citenamefont
  {Jia}, \citenamefont {Dai}, \citenamefont {Fang}, \citenamefont {Zhang},
  \citenamefont {He}, \citenamefont {Wang}, \citenamefont {Lu}, \citenamefont
  {Ma},\ and\ \citenamefont {Xue}}]{Chang2013}%
  \BibitemOpen
  \bibfield  {author} {\bibinfo {author} {\bibfnamefont {Cui-Zu}\ \bibnamefont
  {Chang}}, \bibinfo {author} {\bibfnamefont {Jinsong}\ \bibnamefont {Zhang}},
  \bibinfo {author} {\bibfnamefont {Xiao}\ \bibnamefont {Feng}}, \bibinfo
  {author} {\bibfnamefont {Jie}\ \bibnamefont {Shen}}, \bibinfo {author}
  {\bibfnamefont {Zuocheng}\ \bibnamefont {Zhang}}, \bibinfo {author}
  {\bibfnamefont {Minghua}\ \bibnamefont {Guo}}, \bibinfo {author}
  {\bibfnamefont {Kang}\ \bibnamefont {Li}}, \bibinfo {author} {\bibfnamefont
  {Yunbo}\ \bibnamefont {Ou}}, \bibinfo {author} {\bibfnamefont {Pang}\
  \bibnamefont {Wei}}, \bibinfo {author} {\bibfnamefont {Li-Li}\ \bibnamefont
  {Wang}}, \bibinfo {author} {\bibfnamefont {Zhong-Qing}\ \bibnamefont {Ji}},
  \bibinfo {author} {\bibfnamefont {Yang}\ \bibnamefont {Feng}}, \bibinfo
  {author} {\bibfnamefont {Shuaihua}\ \bibnamefont {Ji}}, \bibinfo {author}
  {\bibfnamefont {Xi}~\bibnamefont {Chen}}, \bibinfo {author} {\bibfnamefont
  {Jinfeng}\ \bibnamefont {Jia}}, \bibinfo {author} {\bibfnamefont
  {Xi}~\bibnamefont {Dai}}, \bibinfo {author} {\bibfnamefont {Zhong}\
  \bibnamefont {Fang}}, \bibinfo {author} {\bibfnamefont {Shou-Cheng}\
  \bibnamefont {Zhang}}, \bibinfo {author} {\bibfnamefont {Ke}~\bibnamefont
  {He}}, \bibinfo {author} {\bibfnamefont {Yayu}\ \bibnamefont {Wang}},
  \bibinfo {author} {\bibfnamefont {Li}~\bibnamefont {Lu}}, \bibinfo {author}
  {\bibfnamefont {Xu-Cun}\ \bibnamefont {Ma}}, \ and\ \bibinfo {author}
  {\bibfnamefont {Qi-Kun}\ \bibnamefont {Xue}},\ }\bibfield  {title} {\enquote
  {\bibinfo {title} {{{Experimental Observation of the Quantum Anomalous Hall
  Effect in a Magnetic Topological Insulator}}},}\ }\href {\doibase
  10.1126/science.1234414} {\bibfield  {journal} {\bibinfo  {journal}
  {Science}\ }\textbf {\bibinfo {volume} {340}},\ \bibinfo {pages} {167--}
  (\bibinfo {year} {2013})}\BibitemShut {NoStop}%
\bibitem [{\citenamefont {Fu}\ and\ \citenamefont {Kane}(2008)}]{Fu2008}%
  \BibitemOpen
  \bibfield  {author} {\bibinfo {author} {\bibfnamefont {Liang}\ \bibnamefont
  {Fu}}\ and\ \bibinfo {author} {\bibfnamefont {C.~L.}\ \bibnamefont {Kane}},\
  }\bibfield  {title} {\enquote {\bibinfo {title} {{{Superconducting Proximity
  Effect and Majorana Fermions at the Surface of a Topological Insulator}}},}\
  }\href {\doibase 10.1103/PhysRevLett.100.096407} {\bibfield  {journal}
  {\bibinfo  {journal} {Phys. Rev. Lett.}\ }\textbf {\bibinfo {volume} {100}},\
  \bibinfo {pages} {096407} (\bibinfo {year} {2008})}\BibitemShut {NoStop}%
\bibitem [{\citenamefont {Qi}\ \emph {et~al.}(2009)\citenamefont {Qi},
  \citenamefont {Hughes}, \citenamefont {Raghu},\ and\ \citenamefont
  {Zhang}}]{Qi2009}%
  \BibitemOpen
  \bibfield  {author} {\bibinfo {author} {\bibfnamefont {Xiao-Liang}\
  \bibnamefont {Qi}}, \bibinfo {author} {\bibfnamefont {Taylor~L.}\
  \bibnamefont {Hughes}}, \bibinfo {author} {\bibfnamefont {S.}~\bibnamefont
  {Raghu}}, \ and\ \bibinfo {author} {\bibfnamefont {Shou-Cheng}\ \bibnamefont
  {Zhang}},\ }\bibfield  {title} {\enquote {\bibinfo {title}
  {{{Time-Reversal-Invariant Topological Superconductors and Superfluids in Two
  and Three Dimensions}}},}\ }\href {\doibase 10.1103/PhysRevLett.102.187001}
  {\bibfield  {journal} {\bibinfo  {journal} {Phys. Rev. Lett.}\ }\textbf
  {\bibinfo {volume} {102}},\ \bibinfo {pages} {187001} (\bibinfo {year}
  {2009})}\BibitemShut {NoStop}%
\bibitem [{\citenamefont {Nadj-Perge}\ \emph {et~al.}(2014)\citenamefont
  {Nadj-Perge}, \citenamefont {Drozdov}, \citenamefont {Li}, \citenamefont
  {Chen}, \citenamefont {Jeon}, \citenamefont {Seo}, \citenamefont {MacDonald},
  \citenamefont {Bernevig},\ and\ \citenamefont {Yazdani}}]{Nadj-Perge2014}%
  \BibitemOpen
  \bibfield  {author} {\bibinfo {author} {\bibfnamefont {Stevan}\ \bibnamefont
  {Nadj-Perge}}, \bibinfo {author} {\bibfnamefont {Ilya~K.}\ \bibnamefont
  {Drozdov}}, \bibinfo {author} {\bibfnamefont {Jian}\ \bibnamefont {Li}},
  \bibinfo {author} {\bibfnamefont {Hua}\ \bibnamefont {Chen}}, \bibinfo
  {author} {\bibfnamefont {Sangjun}\ \bibnamefont {Jeon}}, \bibinfo {author}
  {\bibfnamefont {Jungpil}\ \bibnamefont {Seo}}, \bibinfo {author}
  {\bibfnamefont {Allan~H.}\ \bibnamefont {MacDonald}}, \bibinfo {author}
  {\bibfnamefont {B.~Andrei}\ \bibnamefont {Bernevig}}, \ and\ \bibinfo
  {author} {\bibfnamefont {Ali}\ \bibnamefont {Yazdani}},\ }\bibfield  {title}
  {\enquote {\bibinfo {title} {{{Observation of Majorana fermions in
  ferromagnetic atomic chains on a superconductor}}},}\ }\href {\doibase
  10.1126/science.1259327} {\bibfield  {journal} {\bibinfo  {journal}
  {Science}\ }\textbf {\bibinfo {volume} {346}},\ \bibinfo {pages} {602--}
  (\bibinfo {year} {2014})}\BibitemShut {NoStop}%
\bibitem [{\citenamefont {Lian}\ \emph {et~al.}(2018)\citenamefont {Lian},
  \citenamefont {Sun}, \citenamefont {Vaezi}, \citenamefont {Qi},\ and\
  \citenamefont {Zhang}}]{Lian2018}%
  \BibitemOpen
  \bibfield  {author} {\bibinfo {author} {\bibfnamefont {Biao}\ \bibnamefont
  {Lian}}, \bibinfo {author} {\bibfnamefont {Xiao-Qi}\ \bibnamefont {Sun}},
  \bibinfo {author} {\bibfnamefont {Abolhassan}\ \bibnamefont {Vaezi}},
  \bibinfo {author} {\bibfnamefont {Xiao-Liang}\ \bibnamefont {Qi}}, \ and\
  \bibinfo {author} {\bibfnamefont {Shou-Cheng}\ \bibnamefont {Zhang}},\
  }\bibfield  {title} {\enquote {\bibinfo {title} {{{Topological quantum
  computation based on chiral Majorana fermions}}},}\ }\href {\doibase
  10.1073/pnas.1810003115} {\bibfield  {journal} {\bibinfo  {journal} {Proc
  Natl Acad Sci USA}\ }\textbf {\bibinfo {volume} {115}},\ \bibinfo {pages}
  {10938--} (\bibinfo {year} {2018})}\BibitemShut {NoStop}%
\bibitem [{\citenamefont {Hasan}\ and\ \citenamefont {Kane}(2010)}]{Hasan2010}%
  \BibitemOpen
  \bibfield  {author} {\bibinfo {author} {\bibfnamefont {M.~Z.}\ \bibnamefont
  {Hasan}}\ and\ \bibinfo {author} {\bibfnamefont {C.~L.}\ \bibnamefont
  {Kane}},\ }\bibfield  {title} {\enquote {\bibinfo {title} {{{Colloquium:
  Topological insulators}}},}\ }\href {\doibase 10.1103/RevModPhys.82.3045}
  {\bibfield  {journal} {\bibinfo  {journal} {Rev. Mod. Phys.}\ }\textbf
  {\bibinfo {volume} {82}},\ \bibinfo {pages} {3045--3067} (\bibinfo {year}
  {2010})}\BibitemShut {NoStop}%
\bibitem [{\citenamefont {Qi}\ and\ \citenamefont {Zhang}(2011)}]{Qi2011}%
  \BibitemOpen
  \bibfield  {author} {\bibinfo {author} {\bibfnamefont {Xiao-Liang}\
  \bibnamefont {Qi}}\ and\ \bibinfo {author} {\bibfnamefont {Shou-Cheng}\
  \bibnamefont {Zhang}},\ }\bibfield  {title} {\enquote {\bibinfo {title}
  {{{Topological insulators and superconductors}}},}\ }\href {\doibase
  10.1103/RevModPhys.83.1057} {\bibfield  {journal} {\bibinfo  {journal} {Rev.
  Mod. Phys.}\ }\textbf {\bibinfo {volume} {83}},\ \bibinfo {pages}
  {1057--1110} (\bibinfo {year} {2011})}\BibitemShut {NoStop}%
\bibitem [{\citenamefont {Dziawa}\ \emph {et~al.}(2012)\citenamefont {Dziawa},
  \citenamefont {Kowalski}, \citenamefont {Dybko}, \citenamefont {Buczko},
  \citenamefont {Szczerbakow}, \citenamefont {Szot}, \citenamefont
  {\'{L}usakowska}, \citenamefont {Balasubramanian}, \citenamefont {Wojek},
  \citenamefont {Berntsen}, \citenamefont {Tjernberg},\ and\ \citenamefont
  {Story}}]{Dziawa2012}%
  \BibitemOpen
  \bibfield  {author} {\bibinfo {author} {\bibfnamefont {P.}~\bibnamefont
  {Dziawa}}, \bibinfo {author} {\bibfnamefont {B.~J.}\ \bibnamefont
  {Kowalski}}, \bibinfo {author} {\bibfnamefont {K.}~\bibnamefont {Dybko}},
  \bibinfo {author} {\bibfnamefont {R.}~\bibnamefont {Buczko}}, \bibinfo
  {author} {\bibfnamefont {A.}~\bibnamefont {Szczerbakow}}, \bibinfo {author}
  {\bibfnamefont {M.}~\bibnamefont {Szot}}, \bibinfo {author} {\bibfnamefont
  {E.}~\bibnamefont {\'{L}usakowska}}, \bibinfo {author} {\bibfnamefont
  {T.}~\bibnamefont {Balasubramanian}}, \bibinfo {author} {\bibfnamefont
  {B.~M.}\ \bibnamefont {Wojek}}, \bibinfo {author} {\bibfnamefont {M.~H.}\
  \bibnamefont {Berntsen}}, \bibinfo {author} {\bibfnamefont {O.}~\bibnamefont
  {Tjernberg}}, \ and\ \bibinfo {author} {\bibfnamefont {T.}~\bibnamefont
  {Story}},\ }\bibfield  {title} {\enquote {\bibinfo {title} {Topological
  crystalline insulator states in
  $\mathrm{Pb}_{1-x}\mathrm{Sn}_x\mathrm{Se}$},}\ }\href
  {https://doi.org/10.1038/nmat3449} {\bibfield  {journal} {\bibinfo  {journal}
  {Nature Materials}\ }\textbf {\bibinfo {volume} {11}},\ \bibinfo {pages}
  {1023} (\bibinfo {year} {2012})}\BibitemShut {NoStop}%
\bibitem [{\citenamefont {Wojek}\ \emph {et~al.}(2015)\citenamefont {Wojek},
  \citenamefont {Berntsen}, \citenamefont {Jonsson}, \citenamefont
  {Szczerbakow}, \citenamefont {Dziawa}, \citenamefont {Kowalski},
  \citenamefont {Story},\ and\ \citenamefont {Tjernberg}}]{Wojek2015}%
  \BibitemOpen
  \bibfield  {author} {\bibinfo {author} {\bibfnamefont {B.~M.}\ \bibnamefont
  {Wojek}}, \bibinfo {author} {\bibfnamefont {M.~H.}\ \bibnamefont {Berntsen}},
  \bibinfo {author} {\bibfnamefont {V.}~\bibnamefont {Jonsson}}, \bibinfo
  {author} {\bibfnamefont {A.}~\bibnamefont {Szczerbakow}}, \bibinfo {author}
  {\bibfnamefont {P.}~\bibnamefont {Dziawa}}, \bibinfo {author} {\bibfnamefont
  {B.~J.}\ \bibnamefont {Kowalski}}, \bibinfo {author} {\bibfnamefont
  {T.}~\bibnamefont {Story}}, \ and\ \bibinfo {author} {\bibfnamefont
  {O.}~\bibnamefont {Tjernberg}},\ }\bibfield  {title} {\enquote {\bibinfo
  {title} {{{Direct observation and temperature control of the surface Dirac
  gap in a topological crystalline insulator}}},}\ }\href {\doibase
  10.1038/ncomms9463} {\bibfield  {journal} {\bibinfo  {journal} {Nature
  Communications}\ }\textbf {\bibinfo {volume} {6}},\ \bibinfo {pages} {8463--}
  (\bibinfo {year} {2015})}\BibitemShut {NoStop}%
\bibitem [{\citenamefont {Xu}\ \emph {et~al.}(2018)\citenamefont {Xu},
  \citenamefont {Zhao}, \citenamefont {Marsik}, \citenamefont {Sheveleva},
  \citenamefont {Lyzwa}, \citenamefont {Dai}, \citenamefont {Chen},
  \citenamefont {Qiu},\ and\ \citenamefont {Bernhard}}]{Xu2018}%
  \BibitemOpen
  \bibfield  {author} {\bibinfo {author} {\bibfnamefont {B.}~\bibnamefont
  {Xu}}, \bibinfo {author} {\bibfnamefont {L.~X.}\ \bibnamefont {Zhao}},
  \bibinfo {author} {\bibfnamefont {P.}~\bibnamefont {Marsik}}, \bibinfo
  {author} {\bibfnamefont {E.}~\bibnamefont {Sheveleva}}, \bibinfo {author}
  {\bibfnamefont {F.}~\bibnamefont {Lyzwa}}, \bibinfo {author} {\bibfnamefont
  {Y.~M.}\ \bibnamefont {Dai}}, \bibinfo {author} {\bibfnamefont {G.~F.}\
  \bibnamefont {Chen}}, \bibinfo {author} {\bibfnamefont {X.~G.}\ \bibnamefont
  {Qiu}}, \ and\ \bibinfo {author} {\bibfnamefont {C.}~\bibnamefont
  {Bernhard}},\ }\bibfield  {title} {\enquote {\bibinfo {title}
  {Temperature-driven topological phase transition and intermediate {D}irac
  semimetal phase in {Z}r{T}e$_5$},}\ }\href@noop {} {\bibfield  {journal}
  {\bibinfo  {journal} {Phys. Rev. Lett.}\ }\textbf {\bibinfo {volume} {121}},\
  \bibinfo {pages} {187401} (\bibinfo {year} {2018})}\BibitemShut {NoStop}%
\bibitem [{\citenamefont {Kadykov}\ \emph {et~al.}(2018)\citenamefont
  {Kadykov}, \citenamefont {Krishtopenko}, \citenamefont {Jouault},
  \citenamefont {Desrat}, \citenamefont {Knap}, \citenamefont {Ruffenach},
  \citenamefont {Consejo}, \citenamefont {Torres}, \citenamefont {Morozov},
  \citenamefont {Mikhailov}, \citenamefont {Dvoretskii},\ and\ \citenamefont
  {Teppe}}]{Kadykov2018}%
  \BibitemOpen
  \bibfield  {author} {\bibinfo {author} {\bibfnamefont {A.~M.}\ \bibnamefont
  {Kadykov}}, \bibinfo {author} {\bibfnamefont {S.~S.}\ \bibnamefont
  {Krishtopenko}}, \bibinfo {author} {\bibfnamefont {B.}~\bibnamefont
  {Jouault}}, \bibinfo {author} {\bibfnamefont {W.}~\bibnamefont {Desrat}},
  \bibinfo {author} {\bibfnamefont {W.}~\bibnamefont {Knap}}, \bibinfo {author}
  {\bibfnamefont {S.}~\bibnamefont {Ruffenach}}, \bibinfo {author}
  {\bibfnamefont {C.}~\bibnamefont {Consejo}}, \bibinfo {author} {\bibfnamefont
  {J.}~\bibnamefont {Torres}}, \bibinfo {author} {\bibfnamefont {S.~V.}\
  \bibnamefont {Morozov}}, \bibinfo {author} {\bibfnamefont {N.~N.}\
  \bibnamefont {Mikhailov}}, \bibinfo {author} {\bibfnamefont {S.~A.}\
  \bibnamefont {Dvoretskii}}, \ and\ \bibinfo {author} {\bibfnamefont
  {F.}~\bibnamefont {Teppe}},\ }\bibfield  {title} {\enquote {\bibinfo {title}
  {{{Temperature-Induced Topological Phase Transition in HgTe Quantum
  Wells}}},}\ }\href {\doibase 10.1103/PhysRevLett.120.086401} {\bibfield
  {journal} {\bibinfo  {journal} {Phys. Rev. Lett.}\ }\textbf {\bibinfo
  {volume} {120}},\ \bibinfo {pages} {086401} (\bibinfo {year}
  {2018})}\BibitemShut {NoStop}%
\bibitem [{\citenamefont {Krizman}\ \emph {et~al.}(2018)\citenamefont
  {Krizman}, \citenamefont {Assaf}, \citenamefont {Phuphachong}, \citenamefont
  {Bauer}, \citenamefont {Springholz}, \citenamefont {de~Vaulchier},\ and\
  \citenamefont {Guldner}}]{Krizman2018}%
  \BibitemOpen
  \bibfield  {author} {\bibinfo {author} {\bibfnamefont {G.}~\bibnamefont
  {Krizman}}, \bibinfo {author} {\bibfnamefont {B.~A.}\ \bibnamefont {Assaf}},
  \bibinfo {author} {\bibfnamefont {T.}~\bibnamefont {Phuphachong}}, \bibinfo
  {author} {\bibfnamefont {G.}~\bibnamefont {Bauer}}, \bibinfo {author}
  {\bibfnamefont {G.}~\bibnamefont {Springholz}}, \bibinfo {author}
  {\bibfnamefont {L.~A.}\ \bibnamefont {de~Vaulchier}}, \ and\ \bibinfo
  {author} {\bibfnamefont {Y.}~\bibnamefont {Guldner}},\ }\bibfield  {title}
  {\enquote {\bibinfo {title} {{{Dirac parameters and topological phase diagram
  of
  $\mathrm{P}{\mathrm{b}}_{1\ensuremath{-}x}\mathrm{S}{\mathrm{n}}_{x}\mathrm{Se}$
  from magnetospectroscopy}}},}\ }\href {\doibase 10.1103/PhysRevB.98.245202}
  {\bibfield  {journal} {\bibinfo  {journal} {Phys. Rev. B}\ }\textbf {\bibinfo
  {volume} {98}},\ \bibinfo {pages} {245202} (\bibinfo {year}
  {2018})}\BibitemShut {NoStop}%
\bibitem [{\citenamefont {Garate}(2013)}]{Garate2013}%
  \BibitemOpen
  \bibfield  {author} {\bibinfo {author} {\bibfnamefont {Ion}\ \bibnamefont
  {Garate}},\ }\bibfield  {title} {\enquote {\bibinfo {title} {Phonon-induced
  topological transitions and crossovers in dirac materials},}\ }\href
  {\doibase 10.1103/PhysRevLett.110.046402} {\bibfield  {journal} {\bibinfo
  {journal} {Phys. Rev. Lett.}\ }\textbf {\bibinfo {volume} {110}},\ \bibinfo
  {pages} {046402} (\bibinfo {year} {2013})}\BibitemShut {NoStop}%
\bibitem [{\citenamefont {Saha}\ and\ \citenamefont {Garate}(2014)}]{Saha2014}%
  \BibitemOpen
  \bibfield  {author} {\bibinfo {author} {\bibfnamefont {Kush}\ \bibnamefont
  {Saha}}\ and\ \bibinfo {author} {\bibfnamefont {Ion}\ \bibnamefont
  {Garate}},\ }\bibfield  {title} {\enquote {\bibinfo {title} {Phonon-induced
  topological insulation},}\ }\href {\doibase 10.1103/PhysRevB.89.205103}
  {\bibfield  {journal} {\bibinfo  {journal} {Phys. Rev. B}\ }\textbf {\bibinfo
  {volume} {89}},\ \bibinfo {pages} {205103} (\bibinfo {year}
  {2014})}\BibitemShut {NoStop}%
\bibitem [{\citenamefont {Kim}\ and\ \citenamefont {Jhi}(2015)}]{Kim2015c}%
  \BibitemOpen
  \bibfield  {author} {\bibinfo {author} {\bibfnamefont {Jinwoong}\
  \bibnamefont {Kim}}\ and\ \bibinfo {author} {\bibfnamefont {Seung-Hoon}\
  \bibnamefont {Jhi}},\ }\bibfield  {title} {\enquote {\bibinfo {title}
  {Topological phase transitions in group {IV}-{VI} semiconductors by
  phonons},}\ }\href {\doibase 10.1103/PhysRevB.92.125142} {\bibfield
  {journal} {\bibinfo  {journal} {Phys. Rev. B}\ }\textbf {\bibinfo {volume}
  {92}},\ \bibinfo {pages} {125142} (\bibinfo {year} {2015})}\BibitemShut
  {NoStop}%
\bibitem [{\citenamefont {Monserrat}\ and\ \citenamefont
  {Vanderbilt}(2016)}]{Monserrat2016}%
  \BibitemOpen
  \bibfield  {author} {\bibinfo {author} {\bibfnamefont {Bartomeu}\
  \bibnamefont {Monserrat}}\ and\ \bibinfo {author} {\bibfnamefont {David}\
  \bibnamefont {Vanderbilt}},\ }\bibfield  {title} {\enquote {\bibinfo {title}
  {Temperature effects in the band structure of topological insulators},}\
  }\href {\doibase 10.1103/PhysRevLett.117.226801} {\bibfield  {journal}
  {\bibinfo  {journal} {Phys. Rev. Lett.}\ }\textbf {\bibinfo {volume} {117}},\
  \bibinfo {pages} {226801} (\bibinfo {year} {2016})}\BibitemShut {NoStop}%
\bibitem [{\citenamefont {Antonius}\ and\ \citenamefont
  {Louie}(2016)}]{Antonius2016}%
  \BibitemOpen
  \bibfield  {author} {\bibinfo {author} {\bibfnamefont {Gabriel}\ \bibnamefont
  {Antonius}}\ and\ \bibinfo {author} {\bibfnamefont {Steven~G.}\ \bibnamefont
  {Louie}},\ }\bibfield  {title} {\enquote {\bibinfo {title}
  {Temperature-induced topological phase transitions: Promoted versus
  suppressed nontrivial topology},}\ }\href {\doibase
  10.1103/PhysRevLett.117.246401} {\bibfield  {journal} {\bibinfo  {journal}
  {Phys. Rev. Lett.}\ }\textbf {\bibinfo {volume} {117}},\ \bibinfo {pages}
  {246401} (\bibinfo {year} {2016})}\BibitemShut {NoStop}%
\bibitem [{\citenamefont {Wang}\ \emph
  {et~al.}(2017{\natexlab{a}})\citenamefont {Wang}, \citenamefont {Jo},
  \citenamefont {Wu}, \citenamefont {Wu}, \citenamefont {Kaminski},
  \citenamefont {Canfield},\ and\ \citenamefont {Johnson}}]{Wang2017g}%
  \BibitemOpen
  \bibfield  {author} {\bibinfo {author} {\bibfnamefont {Lin-Lin}\ \bibnamefont
  {Wang}}, \bibinfo {author} {\bibfnamefont {Na~Hyun}\ \bibnamefont {Jo}},
  \bibinfo {author} {\bibfnamefont {Yun}\ \bibnamefont {Wu}}, \bibinfo {author}
  {\bibfnamefont {QuanSheng}\ \bibnamefont {Wu}}, \bibinfo {author}
  {\bibfnamefont {Adam}\ \bibnamefont {Kaminski}}, \bibinfo {author}
  {\bibfnamefont {Paul~C.}\ \bibnamefont {Canfield}}, \ and\ \bibinfo {author}
  {\bibfnamefont {Duane~D.}\ \bibnamefont {Johnson}},\ }\bibfield  {title}
  {\enquote {\bibinfo {title} {{{Phonon-induced topological transition to a
  type-II Weyl semimetal}}},}\ }\href {\doibase 10.1103/PhysRevB.95.165114}
  {\bibfield  {journal} {\bibinfo  {journal} {Phys. Rev. B}\ }\textbf {\bibinfo
  {volume} {95}},\ \bibinfo {pages} {165114} (\bibinfo {year}
  {2017}{\natexlab{a}})}\BibitemShut {NoStop}%
\bibitem [{\citenamefont {Monserrat}\ and\ \citenamefont
  {Vanderbilt}(2017)}]{bitei_rashba_temp}%
  \BibitemOpen
  \bibfield  {author} {\bibinfo {author} {\bibfnamefont {Bartomeu}\
  \bibnamefont {Monserrat}}\ and\ \bibinfo {author} {\bibfnamefont {David}\
  \bibnamefont {Vanderbilt}},\ }\bibfield  {title} {\enquote {\bibinfo {title}
  {Temperature dependence of the bulk {R}ashba splitting in the bismuth
  tellurohalides},}\ }\href
  {https://link.aps.org/doi/10.1103/PhysRevMaterials.1.054201} {\bibfield
  {journal} {\bibinfo  {journal} {Phys. Rev. Materials}\ }\textbf {\bibinfo
  {volume} {1}},\ \bibinfo {pages} {054201} (\bibinfo {year}
  {2017})}\BibitemShut {NoStop}%
\bibitem [{\citenamefont {M\"{o}ller}\ \emph {et~al.}(2017)\citenamefont
  {M\"{o}ller}, \citenamefont {Sawatzky}, \citenamefont {Franz},\ and\
  \citenamefont {Berciu}}]{moller_typeII_elph}%
  \BibitemOpen
  \bibfield  {author} {\bibinfo {author} {\bibfnamefont {Mirko~M.}\
  \bibnamefont {M\"{o}ller}}, \bibinfo {author} {\bibfnamefont {George~A.}\
  \bibnamefont {Sawatzky}}, \bibinfo {author} {\bibfnamefont {Marcel}\
  \bibnamefont {Franz}}, \ and\ \bibinfo {author} {\bibfnamefont {Mona}\
  \bibnamefont {Berciu}},\ }\bibfield  {title} {\enquote {\bibinfo {title}
  {Type-{II} {D}irac semimetal stabilized by electron-phonon coupling},}\
  }\href {https://doi.org/10.1038/s41467-017-02442-y} {\bibfield  {journal}
  {\bibinfo  {journal} {Nature Commun.}\ }\textbf {\bibinfo {volume} {8}},\
  \bibinfo {pages} {2267} (\bibinfo {year} {2017})}\BibitemShut {NoStop}%
\bibitem [{\citenamefont {Harada}\ \emph {et~al.}(1981)\citenamefont {Harada},
  \citenamefont {Sasa},\ and\ \citenamefont {Uda}}]{pbo2_exp_lattice}%
  \BibitemOpen
  \bibfield  {author} {\bibinfo {author} {\bibfnamefont {H.}~\bibnamefont
  {Harada}}, \bibinfo {author} {\bibfnamefont {Y.}~\bibnamefont {Sasa}}, \ and\
  \bibinfo {author} {\bibfnamefont {M.}~\bibnamefont {Uda}},\ }\bibfield
  {title} {\enquote {\bibinfo {title} {{Crystal data for
  $\beta$-{P}b{O}$_2$}},}\ }\href {https://doi.org/10.1107/S0021889881008959}
  {\bibfield  {journal} {\bibinfo  {journal} {J. Appl. Crystallogr.}\ }\textbf
  {\bibinfo {volume} {14}},\ \bibinfo {pages} {141--142} (\bibinfo {year}
  {1981})}\BibitemShut {NoStop}%
\bibitem [{\citenamefont {Kresse}\ and\ \citenamefont
  {Furthm\"uller}(1996)}]{Kresse1996}%
  \BibitemOpen
  \bibfield  {author} {\bibinfo {author} {\bibfnamefont {G.}~\bibnamefont
  {Kresse}}\ and\ \bibinfo {author} {\bibfnamefont {J.}~\bibnamefont
  {Furthm\"uller}},\ }\bibfield  {title} {\enquote {\bibinfo {title}
  {{Efficient iterative schemes for \textit{ab initio} total-energy
  calculations using a plane-wave basis set}},}\ }\href {\doibase
  10.1103/PhysRevB.54.11169} {\bibfield  {journal} {\bibinfo  {journal} {Phys.
  Rev. B}\ }\textbf {\bibinfo {volume} {54}},\ \bibinfo {pages} {11169--11186}
  (\bibinfo {year} {1996})}\BibitemShut {NoStop}%
\bibitem [{\citenamefont {Heyd}\ \emph {et~al.}(2003)\citenamefont {Heyd},
  \citenamefont {Scuseria},\ and\ \citenamefont {Ernzerhof}}]{HSE1}%
  \BibitemOpen
  \bibfield  {author} {\bibinfo {author} {\bibfnamefont {Jochen}\ \bibnamefont
  {Heyd}}, \bibinfo {author} {\bibfnamefont {Gustavo~E.}\ \bibnamefont
  {Scuseria}}, \ and\ \bibinfo {author} {\bibfnamefont {Matthias}\ \bibnamefont
  {Ernzerhof}},\ }\bibfield  {title} {\enquote {\bibinfo {title} {{Hybrid
  functionals based on a screened Coulomb potential}},}\ }\href {\doibase
  10.1063/1.1564060} {\bibfield  {journal} {\bibinfo  {journal} {J. Chem.
  Phys.}\ }\textbf {\bibinfo {volume} {118}},\ \bibinfo {pages} {8207}
  (\bibinfo {year} {2003})}\BibitemShut {NoStop}%
\bibitem [{\citenamefont {Heyd}\ \emph {et~al.}(2006)\citenamefont {Heyd},
  \citenamefont {Scuseria},\ and\ \citenamefont {Ernzerhof}}]{HSE2}%
  \BibitemOpen
  \bibfield  {author} {\bibinfo {author} {\bibfnamefont {Jochen}\ \bibnamefont
  {Heyd}}, \bibinfo {author} {\bibfnamefont {Gustavo~E.}\ \bibnamefont
  {Scuseria}}, \ and\ \bibinfo {author} {\bibfnamefont {Matthias}\ \bibnamefont
  {Ernzerhof}},\ }\bibfield  {title} {\enquote {\bibinfo {title} {{Erratum:
  “Hybrid functionals based on a screened Coulomb potential” [J. Chem.
  Phys.118, 8207 (2003)]}},}\ }\href {\doibase 10.1063/1.2204597} {\bibfield
  {journal} {\bibinfo  {journal} {J. Chem. Phys.}\ }\textbf {\bibinfo {volume}
  {124}},\ \bibinfo {pages} {219906} (\bibinfo {year} {2006})}\BibitemShut
  {NoStop}%
\bibitem [{\citenamefont {Peralta}\ \emph {et~al.}(2006)\citenamefont
  {Peralta}, \citenamefont {Heyd}, \citenamefont {Scuseria},\ and\
  \citenamefont {Martin}}]{HSE3}%
  \BibitemOpen
  \bibfield  {author} {\bibinfo {author} {\bibfnamefont {Juan~E.}\ \bibnamefont
  {Peralta}}, \bibinfo {author} {\bibfnamefont {Jochen}\ \bibnamefont {Heyd}},
  \bibinfo {author} {\bibfnamefont {Gustavo~E.}\ \bibnamefont {Scuseria}}, \
  and\ \bibinfo {author} {\bibfnamefont {Richard~L.}\ \bibnamefont {Martin}},\
  }\bibfield  {title} {\enquote {\bibinfo {title} {{Spin-orbit splittings and
  energy band gaps calculated with the Heyd-Scuseria-Ernzerhof screened hybrid
  functional}},}\ }\href {\doibase 10.1103/PhysRevB.74.073101} {\bibfield
  {journal} {\bibinfo  {journal} {Phys. Rev. B}\ }\textbf {\bibinfo {volume}
  {74}},\ \bibinfo {pages} {073101} (\bibinfo {year} {2006})}\BibitemShut
  {NoStop}%
\bibitem [{\citenamefont {Perdew}\ \emph {et~al.}(1996)\citenamefont {Perdew},
  \citenamefont {Burke},\ and\ \citenamefont {Ernzerhof}}]{Perdew1996}%
  \BibitemOpen
  \bibfield  {author} {\bibinfo {author} {\bibfnamefont {John~P.}\ \bibnamefont
  {Perdew}}, \bibinfo {author} {\bibfnamefont {Kieron}\ \bibnamefont {Burke}},
  \ and\ \bibinfo {author} {\bibfnamefont {Matthias}\ \bibnamefont
  {Ernzerhof}},\ }\bibfield  {title} {\enquote {\bibinfo {title} {Generalized
  gradient approximation made simple},}\ }\href {\doibase
  10.1103/PhysRevLett.77.3865} {\bibfield  {journal} {\bibinfo  {journal}
  {Phys. Rev. Lett.}\ }\textbf {\bibinfo {volume} {77}},\ \bibinfo {pages}
  {3865--3868} (\bibinfo {year} {1996})}\BibitemShut {NoStop}%
\bibitem [{\citenamefont {Mostofi}\ \emph {et~al.}(2014)\citenamefont
  {Mostofi}, \citenamefont {Yates}, \citenamefont {Pizzi}, \citenamefont {Lee},
  \citenamefont {Souza}, \citenamefont {Vanderbilt},\ and\ \citenamefont
  {Marzari}}]{Mostofi2014}%
  \BibitemOpen
  \bibfield  {author} {\bibinfo {author} {\bibfnamefont {Arash~A.}\
  \bibnamefont {Mostofi}}, \bibinfo {author} {\bibfnamefont {Jonathan~R.}\
  \bibnamefont {Yates}}, \bibinfo {author} {\bibfnamefont {Giovanni}\
  \bibnamefont {Pizzi}}, \bibinfo {author} {\bibfnamefont {Young-Su}\
  \bibnamefont {Lee}}, \bibinfo {author} {\bibfnamefont {Ivo}\ \bibnamefont
  {Souza}}, \bibinfo {author} {\bibfnamefont {David}\ \bibnamefont
  {Vanderbilt}}, \ and\ \bibinfo {author} {\bibfnamefont {Nicola}\ \bibnamefont
  {Marzari}},\ }\bibfield  {title} {\enquote {\bibinfo {title} {{An updated
  version of Wannier90: A tool for obtaining maximally-localised Wannier
  functions}},}\ }\href {\doibase 10.1016/j.cpc.2014.05.003} {\bibfield
  {journal} {\bibinfo  {journal} {Computer Physics Communications}\ }\textbf
  {\bibinfo {volume} {185}},\ \bibinfo {pages} {2309--2310} (\bibinfo {year}
  {2014})}\BibitemShut {NoStop}%
\bibitem [{\citenamefont {Zhang}\ \emph {et~al.}(2010)\citenamefont {Zhang},
  \citenamefont {Yu}, \citenamefont {Zhang}, \citenamefont {Dai},\ and\
  \citenamefont {Fang}}]{Zhang2010a}%
  \BibitemOpen
  \bibfield  {author} {\bibinfo {author} {\bibfnamefont {Wei}\ \bibnamefont
  {Zhang}}, \bibinfo {author} {\bibfnamefont {Rui}\ \bibnamefont {Yu}},
  \bibinfo {author} {\bibfnamefont {Hai-Jun}\ \bibnamefont {Zhang}}, \bibinfo
  {author} {\bibfnamefont {Xi}~\bibnamefont {Dai}}, \ and\ \bibinfo {author}
  {\bibfnamefont {Zhong}\ \bibnamefont {Fang}},\ }\bibfield  {title} {\enquote
  {\bibinfo {title} {{First-principles studies of the three-dimensional strong
  topological insulators Bi$_2$Te$_3$, Bi$_2$Se$_3$ and Sb$_2$Te$_3$}},}\
  }\href {\doibase 10.1088/1367-2630/12/6/065013} {\bibfield  {journal}
  {\bibinfo  {journal} {New Journal of Physics}\ }\textbf {\bibinfo {volume}
  {12}},\ \bibinfo {pages} {065013} (\bibinfo {year} {2010})}\BibitemShut
  {NoStop}%
\bibitem [{\citenamefont {Wu}\ \emph {et~al.}(2018)\citenamefont {Wu},
  \citenamefont {Zhang}, \citenamefont {Song}, \citenamefont {Troyer},\ and\
  \citenamefont {Soluyanov}}]{Wu2018}%
  \BibitemOpen
  \bibfield  {author} {\bibinfo {author} {\bibfnamefont {QuanSheng}\
  \bibnamefont {Wu}}, \bibinfo {author} {\bibfnamefont {ShengNan}\ \bibnamefont
  {Zhang}}, \bibinfo {author} {\bibfnamefont {Hai-Feng}\ \bibnamefont {Song}},
  \bibinfo {author} {\bibfnamefont {Matthias}\ \bibnamefont {Troyer}}, \ and\
  \bibinfo {author} {\bibfnamefont {Alexey~A.}\ \bibnamefont {Soluyanov}},\
  }\bibfield  {title} {\enquote {\bibinfo {title} {{WannierTools: An
  open-source software package for novel topological materials}},}\ }\href
  {\doibase 10.1016/j.cpc.2017.09.033} {\bibfield  {journal} {\bibinfo
  {journal} {Computer Physics Communications}\ }\textbf {\bibinfo {volume}
  {224}},\ \bibinfo {pages} {405--416} (\bibinfo {year} {2018})}\BibitemShut
  {NoStop}%
\bibitem [{\citenamefont {Lloyd-Williams}\ and\ \citenamefont
  {Monserrat}(2015)}]{Lloyd-Williams2015}%
  \BibitemOpen
  \bibfield  {author} {\bibinfo {author} {\bibfnamefont {Jonathan~H.}\
  \bibnamefont {Lloyd-Williams}}\ and\ \bibinfo {author} {\bibfnamefont
  {Bartomeu}\ \bibnamefont {Monserrat}},\ }\bibfield  {title} {\enquote
  {\bibinfo {title} {Lattice dynamics and electron-phonon coupling calculations
  using nondiagonal supercells},}\ }\href {\doibase 10.1103/PhysRevB.92.184301}
  {\bibfield  {journal} {\bibinfo  {journal} {Phys. Rev. B}\ }\textbf {\bibinfo
  {volume} {92}},\ \bibinfo {pages} {184301} (\bibinfo {year}
  {2015})}\BibitemShut {NoStop}%
\bibitem [{\citenamefont {Monserrat}(2018)}]{Monserrat2018}%
  \BibitemOpen
  \bibfield  {author} {\bibinfo {author} {\bibfnamefont {Bartomeu}\
  \bibnamefont {Monserrat}},\ }\bibfield  {title} {\enquote {\bibinfo {title}
  {Electron-phonon coupling from finite differences},}\ }\href {\doibase
  10.1088/1361-648X/aaa737} {\bibfield  {journal} {\bibinfo  {journal} {Journal
  of Physics: Condensed Matter}\ }\textbf {\bibinfo {volume} {30}},\ \bibinfo
  {pages} {083001--} (\bibinfo {year} {2018})}\BibitemShut {NoStop}%
\bibitem [{\citenamefont {Monserrat}(2016{\natexlab{a}})}]{Monserrat2016b}%
  \BibitemOpen
  \bibfield  {author} {\bibinfo {author} {\bibfnamefont {Bartomeu}\
  \bibnamefont {Monserrat}},\ }\bibfield  {title} {\enquote {\bibinfo {title}
  {{{Vibrational averages along thermal lines}}},}\ }\href {\doibase
  10.1103/PhysRevB.93.014302} {\bibfield  {journal} {\bibinfo  {journal} {Phys.
  Rev. B}\ }\textbf {\bibinfo {volume} {93}},\ \bibinfo {pages} {014302}
  (\bibinfo {year} {2016}{\natexlab{a}})}\BibitemShut {NoStop}%
\bibitem [{\citenamefont {Monserrat}(2016{\natexlab{b}})}]{Monserrat2016a}%
  \BibitemOpen
  \bibfield  {author} {\bibinfo {author} {\bibfnamefont {Bartomeu}\
  \bibnamefont {Monserrat}},\ }\bibfield  {title} {\enquote {\bibinfo {title}
  {Correlation effects on electron-phonon coupling in semiconductors: Many-body
  theory along thermal lines},}\ }\href {\doibase 10.1103/PhysRevB.93.100301}
  {\bibfield  {journal} {\bibinfo  {journal} {Phys. Rev. B}\ }\textbf {\bibinfo
  {volume} {93}},\ \bibinfo {pages} {100301(R)} (\bibinfo {year}
  {2016}{\natexlab{b}})}\BibitemShut {NoStop}%
\bibitem [{\citenamefont {Carr}\ and\ \citenamefont
  {Hampson}(1972)}]{Carr1972}%
  \BibitemOpen
  \bibfield  {author} {\bibinfo {author} {\bibfnamefont {J.~P.}\ \bibnamefont
  {Carr}}\ and\ \bibinfo {author} {\bibfnamefont {N.~A.}\ \bibnamefont
  {Hampson}},\ }\bibfield  {title} {\enquote {\bibinfo {title} {Lead dioxide
  electrode},}\ }\href {\doibase 10.1021/cr60280a003} {\bibfield  {journal}
  {\bibinfo  {journal} {Chem. Rev.}\ }\textbf {\bibinfo {volume} {72}},\
  \bibinfo {pages} {679--703} (\bibinfo {year} {1972})}\BibitemShut {NoStop}%
\bibitem [{\citenamefont {Wang}\ \emph
  {et~al.}(2017{\natexlab{b}})\citenamefont {Wang}, \citenamefont {Deng},
  \citenamefont {Jiao}, \citenamefont {Zhou},\ and\ \citenamefont
  {Sun}}]{Wang2017f}%
  \BibitemOpen
  \bibfield  {author} {\bibinfo {author} {\bibfnamefont {Wei}\ \bibnamefont
  {Wang}}, \bibinfo {author} {\bibfnamefont {Linjie}\ \bibnamefont {Deng}},
  \bibinfo {author} {\bibfnamefont {Na}~\bibnamefont {Jiao}}, \bibinfo {author}
  {\bibfnamefont {Pan}\ \bibnamefont {Zhou}}, \ and\ \bibinfo {author}
  {\bibfnamefont {Lizhong}\ \bibnamefont {Sun}},\ }\bibfield  {title} {\enquote
  {\bibinfo {title} {{{Three-Dimensional Dirac Semimetal $\beta$-PbO$_2$}}},}\
  }\href {\doibase 10.1002/pssr.201700271} {\bibfield  {journal} {\bibinfo
  {journal} {Phys. Status Solidi RRL}\ }\textbf {\bibinfo {volume} {11}},\
  \bibinfo {pages} {1700271--} (\bibinfo {year}
  {2017}{\natexlab{b}})}\BibitemShut {NoStop}%
\bibitem [{\citenamefont {Chen}\ \emph {et~al.}(2018)\citenamefont {Chen},
  \citenamefont {Shao}, \citenamefont {Lu}, \citenamefont {Wang}, \citenamefont
  {Wu}, \citenamefont {Sun},\ and\ \citenamefont {Xing}}]{Chen2018a}%
  \BibitemOpen
  \bibfield  {author} {\bibinfo {author} {\bibfnamefont {Tong}\ \bibnamefont
  {Chen}}, \bibinfo {author} {\bibfnamefont {Dexi}\ \bibnamefont {Shao}},
  \bibinfo {author} {\bibfnamefont {Pengchao}\ \bibnamefont {Lu}}, \bibinfo
  {author} {\bibfnamefont {Xiaomeng}\ \bibnamefont {Wang}}, \bibinfo {author}
  {\bibfnamefont {Juefei}\ \bibnamefont {Wu}}, \bibinfo {author} {\bibfnamefont
  {Jian}\ \bibnamefont {Sun}}, \ and\ \bibinfo {author} {\bibfnamefont
  {Dingyu}\ \bibnamefont {Xing}},\ }\bibfield  {title} {\enquote {\bibinfo
  {title} {{{Anharmonic effect driven topological phase transition in PbO$_2$
  predicted by first-principles calculations}}},}\ }\href {\doibase
  10.1103/PhysRevB.98.144105} {\bibfield  {journal} {\bibinfo  {journal} {Phys.
  Rev. B}\ }\textbf {\bibinfo {volume} {98}},\ \bibinfo {pages} {144105}
  (\bibinfo {year} {2018})}\BibitemShut {NoStop}%
\bibitem [{\citenamefont {Burgio}\ \emph {et~al.}(2001)\citenamefont {Burgio},
  \citenamefont {Clark},\ and\ \citenamefont {Firth}}]{Burgio2001}%
  \BibitemOpen
  \bibfield  {author} {\bibinfo {author} {\bibfnamefont {Lucia}\ \bibnamefont
  {Burgio}}, \bibinfo {author} {\bibfnamefont {Robin J.~H.}\ \bibnamefont
  {Clark}}, \ and\ \bibinfo {author} {\bibfnamefont {Steven}\ \bibnamefont
  {Firth}},\ }\bibfield  {title} {\enquote {\bibinfo {title} {{{Raman
  spectroscopy as a means for the identification of plattnerite (PbO$_2$), of
  lead pigments and of their degradation products}}},}\ }\href {\doibase
  10.1039/B008302J} {\bibfield  {journal} {\bibinfo  {journal} {Analyst}\
  }\textbf {\bibinfo {volume} {126}},\ \bibinfo {pages} {222--227} (\bibinfo
  {year} {2001})}\BibitemShut {NoStop}%
\bibitem [{\citenamefont {Scanlon}\ \emph {et~al.}(2011)\citenamefont
  {Scanlon}, \citenamefont {Kehoe}, \citenamefont {Watson}, \citenamefont
  {Jones}, \citenamefont {David}, \citenamefont {Payne}, \citenamefont
  {Egdell}, \citenamefont {Edwards},\ and\ \citenamefont
  {Walsh}}]{Scanlon2011}%
  \BibitemOpen
  \bibfield  {author} {\bibinfo {author} {\bibfnamefont {David~O.}\
  \bibnamefont {Scanlon}}, \bibinfo {author} {\bibfnamefont {Aoife~B.}\
  \bibnamefont {Kehoe}}, \bibinfo {author} {\bibfnamefont {Graeme~W.}\
  \bibnamefont {Watson}}, \bibinfo {author} {\bibfnamefont {Martin~O.}\
  \bibnamefont {Jones}}, \bibinfo {author} {\bibfnamefont {William I.~F.}\
  \bibnamefont {David}}, \bibinfo {author} {\bibfnamefont {David~J.}\
  \bibnamefont {Payne}}, \bibinfo {author} {\bibfnamefont {Russell~G.}\
  \bibnamefont {Egdell}}, \bibinfo {author} {\bibfnamefont {Peter~P.}\
  \bibnamefont {Edwards}}, \ and\ \bibinfo {author} {\bibfnamefont {Aron}\
  \bibnamefont {Walsh}},\ }\bibfield  {title} {\enquote {\bibinfo {title}
  {{{Nature of the Band Gap and Origin of the Conductivity of
  ${\mathrm{PbO}}_{2}$ Revealed by Theory and Experiment}}},}\ }\href {\doibase
  10.1103/PhysRevLett.107.246402} {\bibfield  {journal} {\bibinfo  {journal}
  {Phys. Rev. Lett.}\ }\textbf {\bibinfo {volume} {107}},\ \bibinfo {pages}
  {246402} (\bibinfo {year} {2011})}\BibitemShut {NoStop}%
\bibitem [{Sup()}]{Supplemental_Material_PbO2}%
  \BibitemOpen
  \href@noop {} {}\bibinfo {note} {See Supplemental Material at [URL will be
  inserted by the production group] for the ground state properties, band
  structure and finite temperature analysis of $\beta$-PbO$_2$, which includes
  Refs.~\cite{Wang2017e,Payne2007,Marrazzo2018,Fang2016,Liu2017,Pezzini2017,Shishkin2006,Shishkin2007,Fuchs2007,Dove1993,Monserrat2014}}\BibitemShut
  {NoStop}%
\bibitem [{\citenamefont {Payne}\ \emph {et~al.}(2009)\citenamefont {Payne},
  \citenamefont {Paolicelli}, \citenamefont {Offi}, \citenamefont {Panaccione},
  \citenamefont {Lacovig}, \citenamefont {Beamson}, \citenamefont {Fondacaro},
  \citenamefont {Monaco}, \citenamefont {Vanko},\ and\ \citenamefont
  {Egdell}}]{Payne2009}%
  \BibitemOpen
  \bibfield  {author} {\bibinfo {author} {\bibfnamefont {D.J.}\ \bibnamefont
  {Payne}}, \bibinfo {author} {\bibfnamefont {G.}~\bibnamefont {Paolicelli}},
  \bibinfo {author} {\bibfnamefont {F.}~\bibnamefont {Offi}}, \bibinfo {author}
  {\bibfnamefont {G.}~\bibnamefont {Panaccione}}, \bibinfo {author}
  {\bibfnamefont {P.}~\bibnamefont {Lacovig}}, \bibinfo {author} {\bibfnamefont
  {G.}~\bibnamefont {Beamson}}, \bibinfo {author} {\bibfnamefont
  {A.}~\bibnamefont {Fondacaro}}, \bibinfo {author} {\bibfnamefont
  {G.}~\bibnamefont {Monaco}}, \bibinfo {author} {\bibfnamefont
  {G.}~\bibnamefont {Vanko}}, \ and\ \bibinfo {author} {\bibfnamefont {R.G.}\
  \bibnamefont {Egdell}},\ }\bibfield  {title} {\enquote {\bibinfo {title} {{{A
  study of core and valence levels in $\beta$-PbO$_2$ by hard X-ray
  photoemission}}},}\ }\href {\doibase 10.1016/j.elspec.2008.10.002} {\bibfield
   {journal} {\bibinfo  {journal} {Journal of Electron Spectroscopy and Related
  Phenomena}\ }\textbf {\bibinfo {volume} {169}},\ \bibinfo {pages} {26--34}
  (\bibinfo {year} {2009})}\BibitemShut {NoStop}%
\bibitem [{\citenamefont {Ma}\ \emph {et~al.}(2016)\citenamefont {Ma},
  \citenamefont {Jiao}, \citenamefont {Gao}, \citenamefont {Gu}, \citenamefont
  {Bilic}, \citenamefont {Sanvito},\ and\ \citenamefont {Du}}]{Ma2016a}%
  \BibitemOpen
  \bibfield  {author} {\bibinfo {author} {\bibfnamefont {Fengxian}\
  \bibnamefont {Ma}}, \bibinfo {author} {\bibfnamefont {Yalong}\ \bibnamefont
  {Jiao}}, \bibinfo {author} {\bibfnamefont {Guoping}\ \bibnamefont {Gao}},
  \bibinfo {author} {\bibfnamefont {Yuantong}\ \bibnamefont {Gu}}, \bibinfo
  {author} {\bibfnamefont {Ante}\ \bibnamefont {Bilic}}, \bibinfo {author}
  {\bibfnamefont {Stefano}\ \bibnamefont {Sanvito}}, \ and\ \bibinfo {author}
  {\bibfnamefont {Aijun}\ \bibnamefont {Du}},\ }\bibfield  {title} {\enquote
  {\bibinfo {title} {{{Substantial Band-Gap Tuning and a Strain-Controlled
  Semiconductor to Gapless/Band-Inverted Semimetal Transition in Rutile
  Lead/Stannic Dioxide}}},}\ }\href {\doibase 10.1021/acsami.6b09967}
  {\bibfield  {journal} {\bibinfo  {journal} {ACS Appl. Mater. Interfaces}\
  }\textbf {\bibinfo {volume} {8}},\ \bibinfo {pages} {25667--25673} (\bibinfo
  {year} {2016})}\BibitemShut {NoStop}%
\bibitem [{\citenamefont {Burkov}\ \emph {et~al.}(2011)\citenamefont {Burkov},
  \citenamefont {Hook},\ and\ \citenamefont {Balents}}]{Burkov2011}%
  \BibitemOpen
  \bibfield  {author} {\bibinfo {author} {\bibfnamefont {A.~A.}\ \bibnamefont
  {Burkov}}, \bibinfo {author} {\bibfnamefont {M.~D.}\ \bibnamefont {Hook}}, \
  and\ \bibinfo {author} {\bibfnamefont {Leon}\ \bibnamefont {Balents}},\
  }\bibfield  {title} {\enquote {\bibinfo {title} {Topological nodal
  semimetals},}\ }\href {\doibase 10.1103/PhysRevB.84.235126} {\bibfield
  {journal} {\bibinfo  {journal} {Phys. Rev. B}\ }\textbf {\bibinfo {volume}
  {84}},\ \bibinfo {pages} {235126} (\bibinfo {year} {2011})}\BibitemShut
  {NoStop}%
\bibitem [{\citenamefont {Weng}\ \emph {et~al.}(2015)\citenamefont {Weng},
  \citenamefont {Liang}, \citenamefont {Xu}, \citenamefont {Yu}, \citenamefont
  {Fang}, \citenamefont {Dai},\ and\ \citenamefont {Kawazoe}}]{Weng2015b}%
  \BibitemOpen
  \bibfield  {author} {\bibinfo {author} {\bibfnamefont {Hongming}\
  \bibnamefont {Weng}}, \bibinfo {author} {\bibfnamefont {Yunye}\ \bibnamefont
  {Liang}}, \bibinfo {author} {\bibfnamefont {Qiunan}\ \bibnamefont {Xu}},
  \bibinfo {author} {\bibfnamefont {Rui}\ \bibnamefont {Yu}}, \bibinfo {author}
  {\bibfnamefont {Zhong}\ \bibnamefont {Fang}}, \bibinfo {author}
  {\bibfnamefont {Xi}~\bibnamefont {Dai}}, \ and\ \bibinfo {author}
  {\bibfnamefont {Yoshiyuki}\ \bibnamefont {Kawazoe}},\ }\bibfield  {title}
  {\enquote {\bibinfo {title} {Topological node-line semimetal in
  three-dimensional graphene networks},}\ }\href {\doibase
  10.1103/PhysRevB.92.045108} {\bibfield  {journal} {\bibinfo  {journal} {Phys.
  Rev. B}\ }\textbf {\bibinfo {volume} {92}},\ \bibinfo {pages} {045108}
  (\bibinfo {year} {2015})}\BibitemShut {NoStop}%
\bibitem [{\citenamefont {Bian}\ \emph {et~al.}(2016)\citenamefont {Bian},
  \citenamefont {Chang}, \citenamefont {Sankar}, \citenamefont {Xu},
  \citenamefont {Zheng}, \citenamefont {Neupert}, \citenamefont {Chiu},
  \citenamefont {Huang}, \citenamefont {Chang}, \citenamefont {Belopolski},
  \citenamefont {Sanchez}, \citenamefont {Neupane}, \citenamefont {Alidoust},
  \citenamefont {Liu}, \citenamefont {Wang}, \citenamefont {Lee}, \citenamefont
  {Jeng}, \citenamefont {Zhang}, \citenamefont {Yuan}, \citenamefont {Jia},
  \citenamefont {Bansil}, \citenamefont {Chou}, \citenamefont {Lin},\ and\
  \citenamefont {Hasan}}]{Bian2016}%
  \BibitemOpen
  \bibfield  {author} {\bibinfo {author} {\bibfnamefont {Guang}\ \bibnamefont
  {Bian}}, \bibinfo {author} {\bibfnamefont {Tay-Rong}\ \bibnamefont {Chang}},
  \bibinfo {author} {\bibfnamefont {Raman}\ \bibnamefont {Sankar}}, \bibinfo
  {author} {\bibfnamefont {Su-Yang}\ \bibnamefont {Xu}}, \bibinfo {author}
  {\bibfnamefont {Hao}\ \bibnamefont {Zheng}}, \bibinfo {author} {\bibfnamefont
  {Titus}\ \bibnamefont {Neupert}}, \bibinfo {author} {\bibfnamefont
  {Ching-Kai}\ \bibnamefont {Chiu}}, \bibinfo {author} {\bibfnamefont
  {Shin-Ming}\ \bibnamefont {Huang}}, \bibinfo {author} {\bibfnamefont
  {Guoqing}\ \bibnamefont {Chang}}, \bibinfo {author} {\bibfnamefont {Ilya}\
  \bibnamefont {Belopolski}}, \bibinfo {author} {\bibfnamefont {Daniel~S.}\
  \bibnamefont {Sanchez}}, \bibinfo {author} {\bibfnamefont {Madhab}\
  \bibnamefont {Neupane}}, \bibinfo {author} {\bibfnamefont {Nasser}\
  \bibnamefont {Alidoust}}, \bibinfo {author} {\bibfnamefont {Chang}\
  \bibnamefont {Liu}}, \bibinfo {author} {\bibfnamefont {BaoKai}\ \bibnamefont
  {Wang}}, \bibinfo {author} {\bibfnamefont {Chi-Cheng}\ \bibnamefont {Lee}},
  \bibinfo {author} {\bibfnamefont {Horng-Tay}\ \bibnamefont {Jeng}}, \bibinfo
  {author} {\bibfnamefont {Chenglong}\ \bibnamefont {Zhang}}, \bibinfo {author}
  {\bibfnamefont {Zhujun}\ \bibnamefont {Yuan}}, \bibinfo {author}
  {\bibfnamefont {Shuang}\ \bibnamefont {Jia}}, \bibinfo {author}
  {\bibfnamefont {Arun}\ \bibnamefont {Bansil}}, \bibinfo {author}
  {\bibfnamefont {Fangcheng}\ \bibnamefont {Chou}}, \bibinfo {author}
  {\bibfnamefont {Hsin}\ \bibnamefont {Lin}}, \ and\ \bibinfo {author}
  {\bibfnamefont {M.~Zahid}\ \bibnamefont {Hasan}},\ }\bibfield  {title}
  {\enquote {\bibinfo {title} {{{Topological nodal-line fermions in spin-orbit
  metal PbTaSe$_2$}}},}\ }\href {\doibase 10.1038/ncomms10556} {\bibfield
  {journal} {\bibinfo  {journal} {Nature Communications}\ }\textbf {\bibinfo
  {volume} {7}},\ \bibinfo {pages} {10556} (\bibinfo {year}
  {2016})}\BibitemShut {NoStop}%
\bibitem [{\citenamefont {Kopnin}\ \emph {et~al.}(2011)\citenamefont {Kopnin},
  \citenamefont {Heikkil\"a},\ and\ \citenamefont {Volovik}}]{Kopnin2011}%
  \BibitemOpen
  \bibfield  {author} {\bibinfo {author} {\bibfnamefont {N.~B.}\ \bibnamefont
  {Kopnin}}, \bibinfo {author} {\bibfnamefont {T.~T.}\ \bibnamefont
  {Heikkil\"a}}, \ and\ \bibinfo {author} {\bibfnamefont {G.~E.}\ \bibnamefont
  {Volovik}},\ }\bibfield  {title} {\enquote {\bibinfo {title}
  {High-temperature surface superconductivity in topological flat-band
  systems},}\ }\href {\doibase 10.1103/PhysRevB.83.220503} {\bibfield
  {journal} {\bibinfo  {journal} {Phys. Rev. B}\ }\textbf {\bibinfo {volume}
  {83}},\ \bibinfo {pages} {220503(R)} (\bibinfo {year} {2011})}\BibitemShut
  {NoStop}%
\bibitem [{\citenamefont {Kim}\ \emph {et~al.}(2019)\citenamefont {Kim},
  \citenamefont {Yang},\ and\ \citenamefont {Kim}}]{Kim2019}%
  \BibitemOpen
  \bibfield  {author} {\bibinfo {author} {\bibfnamefont {Rokyeon}\ \bibnamefont
  {Kim}}, \bibinfo {author} {\bibfnamefont {Bohm-Jung}\ \bibnamefont {Yang}}, \
  and\ \bibinfo {author} {\bibfnamefont {Choong~H.}\ \bibnamefont {Kim}},\
  }\bibfield  {title} {\enquote {\bibinfo {title} {{{Crystalline topological
  Dirac semimetal phase in rutile structure
  ${\ensuremath{\beta}}^{\ensuremath{'}}\text{\ensuremath{-}}{\mathrm{PtO}}_{2}$}}},}\
  }\href {\doibase 10.1103/PhysRevB.99.045130} {\bibfield  {journal} {\bibinfo
  {journal} {Phys. Rev. B}\ }\textbf {\bibinfo {volume} {99}},\ \bibinfo
  {pages} {045130} (\bibinfo {year} {2019})}\BibitemShut {NoStop}%
\bibitem [{\citenamefont {Wittkamper}\ \emph {et~al.}(2017)\citenamefont
  {Wittkamper}, \citenamefont {Xu}, \citenamefont {Kombaiah}, \citenamefont
  {Ram}, \citenamefont {De~Graef}, \citenamefont {Kitchin}, \citenamefont
  {Rohrer},\ and\ \citenamefont {Salvador}}]{sno2_epitaxial_growth}%
  \BibitemOpen
  \bibfield  {author} {\bibinfo {author} {\bibfnamefont {Julia}\ \bibnamefont
  {Wittkamper}}, \bibinfo {author} {\bibfnamefont {Zhongnan}\ \bibnamefont
  {Xu}}, \bibinfo {author} {\bibfnamefont {Boopathy}\ \bibnamefont {Kombaiah}},
  \bibinfo {author} {\bibfnamefont {Farangis}\ \bibnamefont {Ram}}, \bibinfo
  {author} {\bibfnamefont {Marc}\ \bibnamefont {De~Graef}}, \bibinfo {author}
  {\bibfnamefont {John~R.}\ \bibnamefont {Kitchin}}, \bibinfo {author}
  {\bibfnamefont {Gregory~S.}\ \bibnamefont {Rohrer}}, \ and\ \bibinfo {author}
  {\bibfnamefont {Paul~A.}\ \bibnamefont {Salvador}},\ }\bibfield  {title}
  {\enquote {\bibinfo {title} {Competitive growth of scrutinyite
  ($\alpha$-{P}b{O}$_2$) and rutile polymorphs of {S}n{O}$_2$ on all
  orientations of columbite {C}o{N}b$_2${O}$_6$ substrates},}\ }\href
  {https://doi.org/10.1021/acs.cgd.7b00569} {\bibfield  {journal} {\bibinfo
  {journal} {Cryst. Growth Des.}\ }\textbf {\bibinfo {volume} {17}},\ \bibinfo
  {pages} {3929--3939} (\bibinfo {year} {2017})}\BibitemShut {NoStop}%
\bibitem [{\citenamefont {Chen}\ \emph {et~al.}(2009)\citenamefont {Chen},
  \citenamefont {Analytis}, \citenamefont {Chu}, \citenamefont {Liu},
  \citenamefont {Mo}, \citenamefont {Qi}, \citenamefont {Zhang}, \citenamefont
  {Lu}, \citenamefont {Dai}, \citenamefont {Fang}, \citenamefont {Zhang},
  \citenamefont {Fisher}, \citenamefont {Hussain},\ and\ \citenamefont
  {Shen}}]{Chen2009_bi2te3}%
  \BibitemOpen
  \bibfield  {author} {\bibinfo {author} {\bibfnamefont {Y.~L.}\ \bibnamefont
  {Chen}}, \bibinfo {author} {\bibfnamefont {J.~G.}\ \bibnamefont {Analytis}},
  \bibinfo {author} {\bibfnamefont {J.-H.}\ \bibnamefont {Chu}}, \bibinfo
  {author} {\bibfnamefont {Z.~K.}\ \bibnamefont {Liu}}, \bibinfo {author}
  {\bibfnamefont {S.-K.}\ \bibnamefont {Mo}}, \bibinfo {author} {\bibfnamefont
  {X.~L.}\ \bibnamefont {Qi}}, \bibinfo {author} {\bibfnamefont {H.~J.}\
  \bibnamefont {Zhang}}, \bibinfo {author} {\bibfnamefont {D.~H.}\ \bibnamefont
  {Lu}}, \bibinfo {author} {\bibfnamefont {X.}~\bibnamefont {Dai}}, \bibinfo
  {author} {\bibfnamefont {Z.}~\bibnamefont {Fang}}, \bibinfo {author}
  {\bibfnamefont {S.~C.}\ \bibnamefont {Zhang}}, \bibinfo {author}
  {\bibfnamefont {I.~R.}\ \bibnamefont {Fisher}}, \bibinfo {author}
  {\bibfnamefont {Z.}~\bibnamefont {Hussain}}, \ and\ \bibinfo {author}
  {\bibfnamefont {Z.-X.}\ \bibnamefont {Shen}},\ }\bibfield  {title} {\enquote
  {\bibinfo {title} {Experimental realization of a three-dimensional
  topological insulator, bi$_2$te$_3$},}\ }\href
  {https://science.sciencemag.org/content/325/5937/178} {\bibfield  {journal}
  {\bibinfo  {journal} {Science}\ }\textbf {\bibinfo {volume} {325}},\ \bibinfo
  {pages} {178--181} (\bibinfo {year} {2009})}\BibitemShut {NoStop}%
\bibitem [{\citenamefont {Lou}\ \emph {et~al.}(2018)\citenamefont {Lou},
  \citenamefont {Guo}, \citenamefont {Li}, \citenamefont {Wang}, \citenamefont
  {Liu}, \citenamefont {Sun}, \citenamefont {Li}, \citenamefont {Wu},
  \citenamefont {Wang}, \citenamefont {Sun}, \citenamefont {Shen},
  \citenamefont {Huang}, \citenamefont {Liu}, \citenamefont {Lu}, \citenamefont
  {Lei}, \citenamefont {Ding},\ and\ \citenamefont {Wang}}]{Lou2018_zrb2}%
  \BibitemOpen
  \bibfield  {author} {\bibinfo {author} {\bibfnamefont {Rui}\ \bibnamefont
  {Lou}}, \bibinfo {author} {\bibfnamefont {Pengjie}\ \bibnamefont {Guo}},
  \bibinfo {author} {\bibfnamefont {Man}\ \bibnamefont {Li}}, \bibinfo {author}
  {\bibfnamefont {Qi}~\bibnamefont {Wang}}, \bibinfo {author} {\bibfnamefont
  {Zhonghao}\ \bibnamefont {Liu}}, \bibinfo {author} {\bibfnamefont {Shanshan}\
  \bibnamefont {Sun}}, \bibinfo {author} {\bibfnamefont {Chenghe}\ \bibnamefont
  {Li}}, \bibinfo {author} {\bibfnamefont {Xuchuan}\ \bibnamefont {Wu}},
  \bibinfo {author} {\bibfnamefont {Zilu}\ \bibnamefont {Wang}}, \bibinfo
  {author} {\bibfnamefont {Zhe}\ \bibnamefont {Sun}}, \bibinfo {author}
  {\bibfnamefont {Dawei}\ \bibnamefont {Shen}}, \bibinfo {author}
  {\bibfnamefont {Yaobo}\ \bibnamefont {Huang}}, \bibinfo {author}
  {\bibfnamefont {Kai}\ \bibnamefont {Liu}}, \bibinfo {author} {\bibfnamefont
  {Zhong-Yi}\ \bibnamefont {Lu}}, \bibinfo {author} {\bibfnamefont {Hechang}\
  \bibnamefont {Lei}}, \bibinfo {author} {\bibfnamefont {Hong}\ \bibnamefont
  {Ding}}, \ and\ \bibinfo {author} {\bibfnamefont {Shancai}\ \bibnamefont
  {Wang}},\ }\bibfield  {title} {\enquote {\bibinfo {title} {Experimental
  observation of bulk nodal lines and electronic surface states in zrb$_2$},}\
  }\href {https://doi.org/10.1038/s41535-018-0121-4} {\bibfield  {journal}
  {\bibinfo  {journal} {npj Quantum Materials}\ }\textbf {\bibinfo {volume}
  {3}},\ \bibinfo {pages} {43} (\bibinfo {year} {2018})}\BibitemShut {NoStop}%
\bibitem [{\citenamefont {Wang}\ and\ \citenamefont {Wang}(2017)}]{Wang2017e}%
  \BibitemOpen
  \bibfield  {author} {\bibinfo {author} {\bibfnamefont {Zhenwei}\ \bibnamefont
  {Wang}}\ and\ \bibinfo {author} {\bibfnamefont {Guangtao}\ \bibnamefont
  {Wang}},\ }\bibfield  {title} {\enquote {\bibinfo {title} {A new strongly
  topological node-line semimetal $\beta$-{P}b{O}$_2$},}\ }\href {\doibase
  10.1016/j.physleta.2017.06.041} {\bibfield  {journal} {\bibinfo  {journal}
  {Physics Letters A}\ }\textbf {\bibinfo {volume} {381}},\ \bibinfo {pages}
  {2856--2859} (\bibinfo {year} {2017})}\BibitemShut {NoStop}%
\bibitem [{\citenamefont {Payne}\ \emph {et~al.}(2007)\citenamefont {Payne},
  \citenamefont {Egdell}, \citenamefont {Law}, \citenamefont {Glans},
  \citenamefont {Learmonth}, \citenamefont {Smith}, \citenamefont {Guo},
  \citenamefont {Walsh},\ and\ \citenamefont {Watson}}]{Payne2007}%
  \BibitemOpen
  \bibfield  {author} {\bibinfo {author} {\bibfnamefont {David~J.}\
  \bibnamefont {Payne}}, \bibinfo {author} {\bibfnamefont {Russell~G.}\
  \bibnamefont {Egdell}}, \bibinfo {author} {\bibfnamefont {Danny S.~L.}\
  \bibnamefont {Law}}, \bibinfo {author} {\bibfnamefont {Per-Anders}\
  \bibnamefont {Glans}}, \bibinfo {author} {\bibfnamefont {Timothy}\
  \bibnamefont {Learmonth}}, \bibinfo {author} {\bibfnamefont {Kevin~E.}\
  \bibnamefont {Smith}}, \bibinfo {author} {\bibfnamefont {Jinghua}\
  \bibnamefont {Guo}}, \bibinfo {author} {\bibfnamefont {Aron}\ \bibnamefont
  {Walsh}}, \ and\ \bibinfo {author} {\bibfnamefont {Graeme~W.}\ \bibnamefont
  {Watson}},\ }\bibfield  {title} {\enquote {\bibinfo {title} {{{Experimental
  and theoretical study of the electronic structures of $\alpha$-PbO and
  $\alpha$-PbO$_2$}}},}\ }\href {\doibase 10.1039/B612323F} {\bibfield
  {journal} {\bibinfo  {journal} {J. Mater. Chem.}\ }\textbf {\bibinfo {volume}
  {17}},\ \bibinfo {pages} {267--277} (\bibinfo {year} {2007})}\BibitemShut
  {NoStop}%
\bibitem [{\citenamefont {Marrazzo}\ \emph {et~al.}(2018)\citenamefont
  {Marrazzo}, \citenamefont {Gibertini}, \citenamefont {Campi}, \citenamefont
  {Mounet},\ and\ \citenamefont {Marzari}}]{Marrazzo2018}%
  \BibitemOpen
  \bibfield  {author} {\bibinfo {author} {\bibfnamefont {Antimo}\ \bibnamefont
  {Marrazzo}}, \bibinfo {author} {\bibfnamefont {Marco}\ \bibnamefont
  {Gibertini}}, \bibinfo {author} {\bibfnamefont {Davide}\ \bibnamefont
  {Campi}}, \bibinfo {author} {\bibfnamefont {Nicolas}\ \bibnamefont {Mounet}},
  \ and\ \bibinfo {author} {\bibfnamefont {Nicola}\ \bibnamefont {Marzari}},\
  }\bibfield  {title} {\enquote {\bibinfo {title} {{{Prediction of a Large-Gap
  and Switchable Kane-Mele Quantum Spin Hall Insulator}}},}\ }\href {\doibase
  10.1103/PhysRevLett.120.117701} {\bibfield  {journal} {\bibinfo  {journal}
  {Phys. Rev. Lett.}\ }\textbf {\bibinfo {volume} {120}},\ \bibinfo {pages}
  {117701} (\bibinfo {year} {2018})}\BibitemShut {NoStop}%
\bibitem [{\citenamefont {Fang}\ \emph {et~al.}(2016)\citenamefont {Fang},
  \citenamefont {Weng}, \citenamefont {Dai},\ and\ \citenamefont
  {Fang}}]{Fang2016}%
  \BibitemOpen
  \bibfield  {author} {\bibinfo {author} {\bibfnamefont {Chen}\ \bibnamefont
  {Fang}}, \bibinfo {author} {\bibfnamefont {Hongming}\ \bibnamefont {Weng}},
  \bibinfo {author} {\bibfnamefont {Xi}~\bibnamefont {Dai}}, \ and\ \bibinfo
  {author} {\bibfnamefont {Zhong}\ \bibnamefont {Fang}},\ }\bibfield  {title}
  {\enquote {\bibinfo {title} {Topological nodal line semimetals},}\ }\href
  {\doibase 10.1088/1674-1056/25/11/117106} {\bibfield  {journal} {\bibinfo
  {journal} {Chinese Physics B}\ }\textbf {\bibinfo {volume} {25}},\ \bibinfo
  {pages} {117106--} (\bibinfo {year} {2016})}\BibitemShut {NoStop}%
\bibitem [{\citenamefont {Liu}\ and\ \citenamefont {Balents}(2017)}]{Liu2017}%
  \BibitemOpen
  \bibfield  {author} {\bibinfo {author} {\bibfnamefont {Jianpeng}\
  \bibnamefont {Liu}}\ and\ \bibinfo {author} {\bibfnamefont {Leon}\
  \bibnamefont {Balents}},\ }\bibfield  {title} {\enquote {\bibinfo {title}
  {Correlation effects and quantum oscillations in topological nodal-loop
  semimetals},}\ }\href {\doibase 10.1103/PhysRevB.95.075426} {\bibfield
  {journal} {\bibinfo  {journal} {Phys. Rev. B}\ }\textbf {\bibinfo {volume}
  {95}},\ \bibinfo {pages} {075426} (\bibinfo {year} {2017})}\BibitemShut
  {NoStop}%
\bibitem [{\citenamefont {Pezzini}\ \emph {et~al.}(2017)\citenamefont
  {Pezzini}, \citenamefont {van Delft}, \citenamefont {Schoop}, \citenamefont
  {Lotsch}, \citenamefont {Carrington}, \citenamefont {Katsnelson},
  \citenamefont {Hussey},\ and\ \citenamefont {Wiedmann}}]{Pezzini2017}%
  \BibitemOpen
  \bibfield  {author} {\bibinfo {author} {\bibfnamefont {S.}~\bibnamefont
  {Pezzini}}, \bibinfo {author} {\bibfnamefont {M.~R.}\ \bibnamefont {van
  Delft}}, \bibinfo {author} {\bibfnamefont {L.~M.}\ \bibnamefont {Schoop}},
  \bibinfo {author} {\bibfnamefont {B.~V.}\ \bibnamefont {Lotsch}}, \bibinfo
  {author} {\bibfnamefont {A.}~\bibnamefont {Carrington}}, \bibinfo {author}
  {\bibfnamefont {M.~I.}\ \bibnamefont {Katsnelson}}, \bibinfo {author}
  {\bibfnamefont {N.~E.}\ \bibnamefont {Hussey}}, \ and\ \bibinfo {author}
  {\bibfnamefont {S.}~\bibnamefont {Wiedmann}},\ }\bibfield  {title} {\enquote
  {\bibinfo {title} {{{Unconventional mass enhancement around the Dirac nodal
  loop in ZrSiS}}},}\ }\href {\doibase 10.1038/nphys4306} {\bibfield  {journal}
  {\bibinfo  {journal} {Nature Physics}\ }\textbf {\bibinfo {volume} {14}},\
  \bibinfo {pages} {178--} (\bibinfo {year} {2017})}\BibitemShut {NoStop}%
\bibitem [{\citenamefont {Shishkin}\ and\ \citenamefont
  {Kresse}(2006)}]{Shishkin2006}%
  \BibitemOpen
  \bibfield  {author} {\bibinfo {author} {\bibfnamefont {M.}~\bibnamefont
  {Shishkin}}\ and\ \bibinfo {author} {\bibfnamefont {G.}~\bibnamefont
  {Kresse}},\ }\bibfield  {title} {\enquote {\bibinfo {title} {Implementation
  and performance of the frequency-dependent $gw$ method within the paw
  framework},}\ }\href {\doibase 10.1103/PhysRevB.74.035101} {\bibfield
  {journal} {\bibinfo  {journal} {Phys. Rev. B}\ }\textbf {\bibinfo {volume}
  {74}},\ \bibinfo {pages} {035101} (\bibinfo {year} {2006})}\BibitemShut
  {NoStop}%
\bibitem [{\citenamefont {Shishkin}\ and\ \citenamefont
  {Kresse}(2007)}]{Shishkin2007}%
  \BibitemOpen
  \bibfield  {author} {\bibinfo {author} {\bibfnamefont {M.}~\bibnamefont
  {Shishkin}}\ and\ \bibinfo {author} {\bibfnamefont {G.}~\bibnamefont
  {Kresse}},\ }\bibfield  {title} {\enquote {\bibinfo {title} {Self-consistent
  $gw$ calculations for semiconductors and insulators},}\ }\href {\doibase
  10.1103/PhysRevB.75.235102} {\bibfield  {journal} {\bibinfo  {journal} {Phys.
  Rev. B}\ }\textbf {\bibinfo {volume} {75}},\ \bibinfo {pages} {235102}
  (\bibinfo {year} {2007})}\BibitemShut {NoStop}%
\bibitem [{\citenamefont {Fuchs}\ \emph {et~al.}(2007)\citenamefont {Fuchs},
  \citenamefont {Furthm\"uller}, \citenamefont {Bechstedt}, \citenamefont
  {Shishkin},\ and\ \citenamefont {Kresse}}]{Fuchs2007}%
  \BibitemOpen
  \bibfield  {author} {\bibinfo {author} {\bibfnamefont {F.}~\bibnamefont
  {Fuchs}}, \bibinfo {author} {\bibfnamefont {J.}~\bibnamefont
  {Furthm\"uller}}, \bibinfo {author} {\bibfnamefont {F.}~\bibnamefont
  {Bechstedt}}, \bibinfo {author} {\bibfnamefont {M.}~\bibnamefont {Shishkin}},
  \ and\ \bibinfo {author} {\bibfnamefont {G.}~\bibnamefont {Kresse}},\
  }\bibfield  {title} {\enquote {\bibinfo {title} {Quasiparticle band structure
  based on a generalized kohn-sham scheme},}\ }\href {\doibase
  10.1103/PhysRevB.76.115109} {\bibfield  {journal} {\bibinfo  {journal} {Phys.
  Rev. B}\ }\textbf {\bibinfo {volume} {76}},\ \bibinfo {pages} {115109}
  (\bibinfo {year} {2007})}\BibitemShut {NoStop}%
\bibitem [{\citenamefont {Dove}(1993)}]{Dove1993}%
  \BibitemOpen
  \bibfield  {author} {\bibinfo {author} {\bibfnamefont {Martin~T.}\
  \bibnamefont {Dove}},\ }\href@noop {} {\emph {\bibinfo {title} {Introduction
  to Lattice Dynamics}}}\ (\bibinfo  {publisher} {Cambridge University Press},\
  \bibinfo {year} {1993})\BibitemShut {NoStop}%
\bibitem [{\citenamefont {Monserrat}\ \emph {et~al.}(2014)\citenamefont
  {Monserrat}, \citenamefont {Conduit},\ and\ \citenamefont
  {Needs}}]{Monserrat2014}%
  \BibitemOpen
  \bibfield  {author} {\bibinfo {author} {\bibfnamefont {Bartomeu}\
  \bibnamefont {Monserrat}}, \bibinfo {author} {\bibfnamefont {G.~J.}\
  \bibnamefont {Conduit}}, \ and\ \bibinfo {author} {\bibfnamefont {R.~J.}\
  \bibnamefont {Needs}},\ }\bibfield  {title} {\enquote {\bibinfo {title}
  {{{Extracting semiconductor band gap zero-point corrections from experimental
  data}}},}\ }\href {\doibase 10.1103/PhysRevB.90.184302} {\bibfield  {journal}
  {\bibinfo  {journal} {Phys. Rev. B}\ }\textbf {\bibinfo {volume} {90}},\
  \bibinfo {pages} {184302} (\bibinfo {year} {2014})}\BibitemShut {NoStop}%
\end{thebibliography}%

\onecolumngrid
\clearpage
\begin{center}
\textbf{\large Supplemental Material for ``Topological semimetallic phase in PbO$_2$ promoted by temperature'' }
\end{center}
\setcounter{equation}{0}
\setcounter{figure}{0}
\setcounter{table}{0}
\setcounter{page}{1}
\makeatletter
\renewcommand{\theequation}{S\arabic{equation}}
\renewcommand{\thefigure}{S\arabic{figure}}
\renewcommand{\bibnumfmt}[1]{[S#1]}
\renewcommand{\citenumfont}[1]{S#1}

\section{Ground state properties of $\beta$-P\lowercase{b}O$_2$}

In this section we provide a detailed analysis of the appropriate level of theory necessary for the correct description of the physical properties of $\beta$-PbO$_2$. Our conclusion is that using density functional theory (DFT) in conjunction with the Heyd-Scuseria-Ernzerhof (HSE06) screened Coulomb hybrid density functional \cite{HSE1,HSE2,HSE3} provides a reliable model for $\beta$-PbO$_2$. In contrast, we show that the widely used semilocal Perdew-Burke-Ernzerhof (PBE) functional is inappropriate for a correct description of this system.

\begin{figure*}[b!]
\centering
\includegraphics[width=\linewidth]{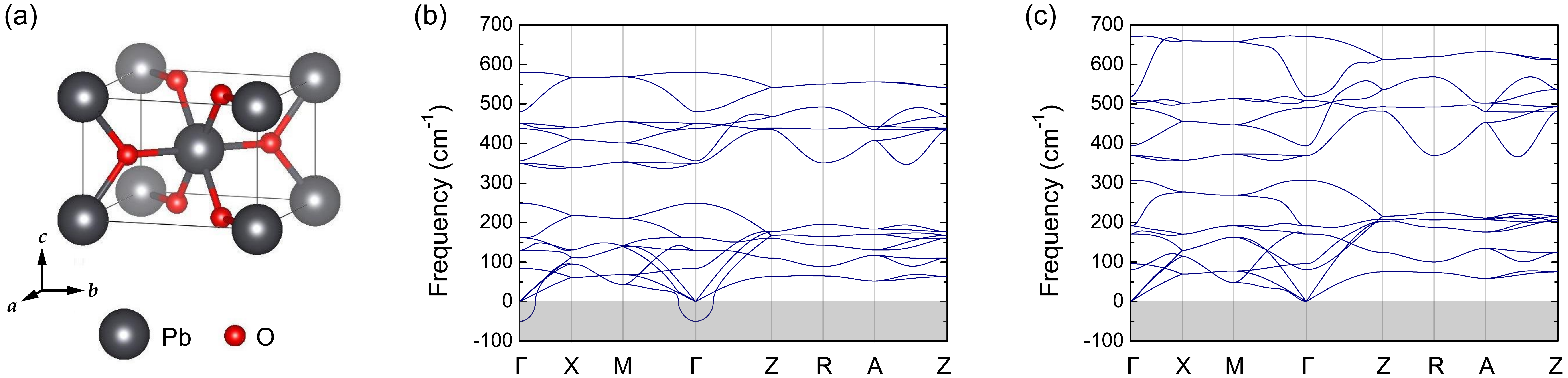}
\caption{(a) Unit cell of the crystal structure of $\beta$-PbO$_2$. Calculated phonon dispersions by using (b) semilocal PBE and (c) hybrid HSE06 functionals.}
\label{f1-si}
\end{figure*}

\subsection*{Crystal structure}

Figure~\ref{f1-si}(a) illustrates the tetragonal structure of the $\beta$-PbO$_2$ phase (space group: $P4_2/mnm$) in which each Pb atom is at the center of an O octahedral cage. Table~\ref{t1} shows a comparison of the lattice parameters of PbO$_2$ calculated with the semilocal PBE and the hybrid HSE06 exchange-correlation functionals, and compared to experimental measurements. The HSE06 lattice parameters are within $0.5$\% of experimental reports \cite{pbo2_exp_lattice}, whereas PBE lattice parameters overestimate the experimental reports by $2.5$\% and $1.9$\% for the \textbf{\textit{a}} and \textbf{\textit{c}} lattice parameters, respectively. These results are in agreement with earlier reports \cite{Scanlon2011}.

\subsection*{Electronic structure}
Despite the prominent role of PbO$_2$ in batteries, basic questions such as whether the $\beta$-PbO$_2$ phase is semiconducting or (semi)metallic remain controversial \cite{Payne2009,Scanlon2011,Wang2017f,Wang2017e,Chen2018a}. 
X-ray photoelectron spectroscopy measurements indicate that $\beta$-PbO$_2$ has a semiconducting gap but the Fermi energy lies within the conduction band \cite{Payne2007,Payne2009}. There is both experimental and theoretical evidence indicating that the partially filled conduction band is due to oxygen deficiency and that $\beta$-PbO$_2$ is an intrinsic semiconductor \cite{Payne2007,Payne2009,Scanlon2011}.

Previous DFT calculations using semilocal exchange-correlation functionals predict that $\beta$-PbO$_2$ is a nodal-line semimetal when spin-orbit coupling is neglected, and becomes a Dirac semimetal in the presence of spin-orbit coupling \cite{Wang2017f,Wang2017e}. In contrast, using hybrid exchange-correlation functionals leads to a semiconducting band gap of about $0.23$\,eV \cite{Scanlon2011}, which is consistent with X-ray photoemission results \cite{Payne2007,Payne2009}, indicating the latter provides a more accurate description of the electronic structure of PbO$_2$. In our calculations, we obtain a semimetallic band structure using the PBE functional and a semiconducting band structure using the HSE06 functional. Using HSE06, we predict a minimum band gap of $0.24$\,eV between the valence band maximum at the R point and the conduction band minimum at the $\Gamma$ point, and a minimum direct band gap at the $\Gamma$ point of $0.34$\,eV. We conclude that the HSE06 functional is necessary for a correct description of the semiconducting nature of $\beta$-PbO$_2$.


\subsection*{Lattice dynamics}

Recent calculations based on a semilocal exchange-correlation functional show that $\beta$-PbO$_2$ is dynamically unstable and that there is a low temperature structure with space group $Pnnm$ that is more stable. These calculations also predict that the experimentally reported $P4_2/mnm$ structure only exists at temperature above 200 K, stabilized by anharmonic lattice dynamics \cite{Chen2018a}.
\begin{table}
  \setlength{\tabcolsep}{6pt} 
\caption{Ground state properties of $\beta$-PbO$_2$ calculated using the semilocal PBE and the hybrid HSE06 functionals. The $a$ and $c$ lattice parameters are shown in \AA, and are compared to the neutron diffraction experiment reported in Ref.\,\cite{pbo2_exp_lattice}. $E_g$, $A_{1g}$, and $B_{1g}$ are the three Raman active modes in cm$^{-1}$, and compared to the experiment reported in Ref.\,\cite{Burgio2001}. We note that the lowest frequency Raman peak at $370$\,cm$^{-1}$ was probably misidentified in the experiments, which reported a value of $424$\,cm$^{-1}$ instead.}
\label{tab:raman}
  \begin{ruledtabular}
\begin{tabular}{c|cccccccc}
 & $a$ & $c$ & $E_g$ & $A_{1g}$ & $B_{1g}$ \\
\hline
PBE        & $5.079$ & $3.448$ & $350.0$ & $480.3$ & $579.5$ \\
HSE06      & $4.974$ & $3.365$ & $369.4$ & $518.4$ & $670.2$ \\
Experiment & $4.961$ & $3.385$ & $370$   & $515$   & $653$ \\
\end{tabular}
\end{ruledtabular}
\label{t1}
\end{table}

Phonon dispersions calculated using semilocal and hybrid exchange-correlation functionals are shown in Fig.\,\ref{f1-si}(b) and (c). There is an imaginary mode at the $\Gamma$ point for the PBE results [Fig.\,\ref{f1-si}(b)], indicating dynamical instability and in good agreement with recent calculations \cite{Chen2018a}. 
However, no low temperature structure has ever been reported experimentally. Using the recently proposed nondiagonal supercells \cite{Lloyd-Williams2015}, we are able to recalculate the phonon dispersion using the computationally expensive HSE06 functional, as shown in Fig.\,\ref{f1-si}(c). Interestingly, the HSE06 phonon dispersion exhibits no imaginary modes, thus predicting that the experimentally observed structure is stable at all temperatures.  This is consistent with the fact that $\beta$-PbO$_2$ is the most stable phase under normal laboratory conditions \cite{Carr1972}. To rationalize these results, we note that the unstable $\Gamma$ point phonon at the PBE level leads to a lower-symmetry structure in which a band gap develops. This Peierls-like mechanism that lowers the density of states at the Fermi level of $\beta$-PbO$_2$ and thus stabilizes the gapped low-symmetry structure at the PBE level is unnecessary at the HSE06 level, because a gap already exists. Analogous physics has been reported in the two-dimensional topological insulator Pt$_2$HgSe$_3$ \cite{Marrazzo2018}, suggesting that good estimates of band gaps are essential for studying topological matter.

To further confirm that HSE06 reproduces the lattice dynamics of PbO$_2$ better than semilocal exchange-correlation functionals, we compare the frequencies of the Raman active phonon modes calculated using PBE and HSE06 with experimental measurements \cite{Burgio2001} in Table \ref{tab:raman}. The HSE06 results show significantly better agreement with experimental measurements.

\subsection*{Ground state properties of $\beta$-PbO$_2$}

The above results indicate that a correct description of $\beta$-PbO$_2$ is only obtained using a hybrid exchange-correlation functional such as HSE06. Consistently with experiment, HSE06 predicts that $\beta$-PbO$_2$ is a small band gap semiconductor with a dynamically stable crystal structure. In the main text, all results have been obtained using the hybrid HSE06 functional.

\begin{figure*}
\centering
\includegraphics[width=\linewidth]{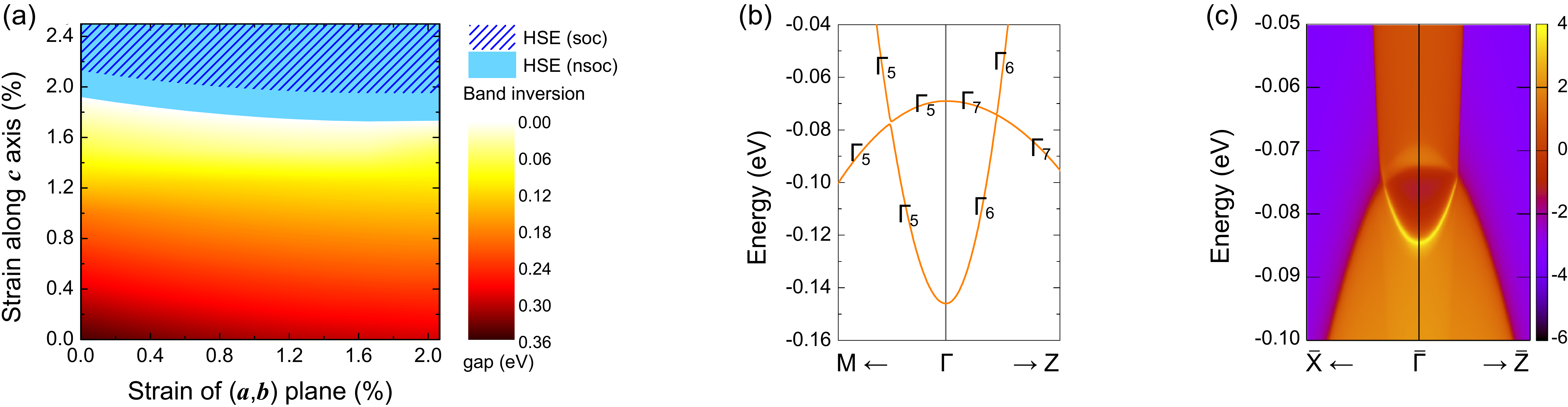}
\caption{(a) Band inversion at the $\Gamma$ point of $\beta$-PbO$_2$ as a function of tensile strain using hybrid HSE06 functional without spin-orbit coupling. The critical strain for band inversion in the presence of spin-orbit coupling is also shown by the blue pattern. (b) Band crossing points of strained PbO$_2$ at the M-$\Gamma$ and $\Gamma$-Z high-symmetry lines in the presence of spin-orbit coupling. (c) The (010) surface states of strained PbO$_2$ in the presence of spin-orbit coupling. A warmer color represents a higher surface contribution.}
\label{f2-si}
\end{figure*}

\section{Band structure}

\subsection{Spin-orbit coupling}
The band edges of $\beta$-PbO$_2$ are comprised of the 6$s$ states of Pb and the 2$p_{xy}$ states of O, and none of these exhibit strong spin-orbit coupling effects. The inclusion of spin-orbit coupling increases the direct band gap of $\beta$-PbO$_2$ by $30.53$\,meV. This implies that the tensile strain required to induce a band inversion increases as well, as shown in Fig.\,\ref{f2-si}(a). However, the weak spin-orbit coupling implies that this increase in the required strain is only of 0.2\%, and in the main text we have neglected the spin-orbit coupling in all presented results.

For spinless systems, the two crossing bands on the (110) and (1$\bar{1}$0) mirror symmetry planes can be labelled by $+1$ and $-1$ mirror symmetry eigenvalues, and are disallowed by mirror symmetry to hybridize with each other, forming two ring-shaped nodal lines on the (110) and (1$\bar{1}$0) planes. Thus two crossing points are observed along the M-$\Gamma$ and $\Gamma$-Z directions in the band structures, which are on the (110) plane. For spinful systems, the mirror eigenvalues on the (110) and (1$\bar{1}$0) planes are $+i$ and $-i$ \cite{Fang2016}. Because $\beta$-PbO$_2$ obeys both time reversal symmetry and inversion symmetry, all the bands are doubly degenerate at every \textbf{k} point. On a plane that is invariant under a mirror operation, a Bloch state and its degenerate partner have different mirror eigenvalues as a result of the commutation relations \cite{Kim2019}. Thus each band has a $+i$ and a $-i$ subband. At the touching point, the subbands with the same eigenvalue repel each other, opening a gap. As a consequence, the inclusion of spin-orbit coupling leads to the disappearence of the ring-shaped nodal lines.

However, the band crossings along the $\Gamma$-Z high-symmetry line are protected even in the presence of spin-orbit coupling by a fourfold screw rotation symmetry $\tilde{C}_{4z}$, which is a fourfold rotation about the \textit{z} axis, followed by a translation by (\textbf{\textit{a}}/2,\textbf{\textit{a}}/2,\textbf{\textit{c}}/2) \cite{Kim2019}. As the band inversion occurs for the two doubly degenerate bands with different screw eigenvalues \{$\tilde{c}_{4z}$(1), $\tilde{c}_{4z}$(2)\} and \{$\tilde{c}_{4z}$(0), $\tilde{c}_{4z}$(3)\}, there is no hybridization between these two bands, and a pair of Dirac nodes are topologically protected along the $k_z$ axis. As shown in Fig.\,\ref{f2-si}(b), the two bands along the $\Gamma$-Z high-symmetry line have different irreducible representations $\Gamma_6$ and $\Gamma_7$. On the other hand, the two bands along the M-$\Gamma$ high-symmetry line have the same irreducible representation $\Gamma_5$, indicating that they can hybridize with each other and become gapped. Along the M-$\Gamma$ high-symmetry line, the spin-orbit coupling induces a gap of $1.05$\,meV. Such a gap can be neglected for temperatures over $12.2$\,K, as the bands are broadened by $k_{\mathrm{B}}T$ and hence merge together above that temperature, similar to three-dimensional graphene networks \cite{Weng2015b}. This implies that, although $\beta$-PbO$_2$ is technically a Dirac semimetal in the band inverted phase, considering thermal broadening means that the physics of interest is that of a nodal line semimetal.

Using the HSE06 band structures, we generate the Wannier functions for the $s$ orbitals of lead and the $p$ orbitals of oxygen. The $(010)$ surface states are calculated using an iterated surface Green's function method \cite{Wu2018}, as shown in Fig.\,\ref{f2-si}(c). It should be noticed that the crossing bands along the $\overline{\textrm{X}}$-$\overline{\Gamma}$ high-symmetry line are slightly gapped, whereas the topological surface states emerge from the projection of bulk Dirac point along the $\overline{\Gamma}$-$\overline{\textrm{Z}}$ high-symmetry line. Because of the weak spin-orbit-coupling effects, the gapping of the Dirac node along the $\overline{\textrm{X}}$-$\overline{\Gamma}$ high-symmetry line is not obvious. Thus the surface states show similar behavior to those of nodal-line semimetals. 
A key feature for the topological nodal-line semimetal is two-dimensional drumhead-like surface states connecting the projected Dirac nodal points, as highlighted by the yellow curve in Fig.\,\ref{f2-si}(c). Such surface states are nestled inside the nodal line with a divergent density of states, and are approximately dispersionless. The slight dispersion in Fig.\,\ref{f2-si}(c) comes from the asymmetry of the inverted bands.

\subsection{Many-body $GW_0$ results}

Due to a divergent density of states arising from surface states in nodal-line semimetals, strong correlation effects may occur \cite{Liu2017,Pezzini2017}, which further increases the uncertainty in accurate calculations of electronic structures. We perform $GW_0$ calculations \cite{Shishkin2006,Shishkin2007,Fuchs2007} using the HSE06 results as a starting point in order to take exchange and correlation interactions of quasiparticles into consideration. In partially self-consistent $GW_0$ calculations, the Green's function $G$ is updated in 8 iterations. The energy cutoff for the response function is set to be 300 eV. A total of 72 (valence and conduction) bands are used with a $\Gamma$-centered \textbf{k}-point sampling of $4\times 4\times 6$. The convergence of our calculations has been checked carefully.

\begin{figure*}
\centering
\includegraphics[width=\linewidth]{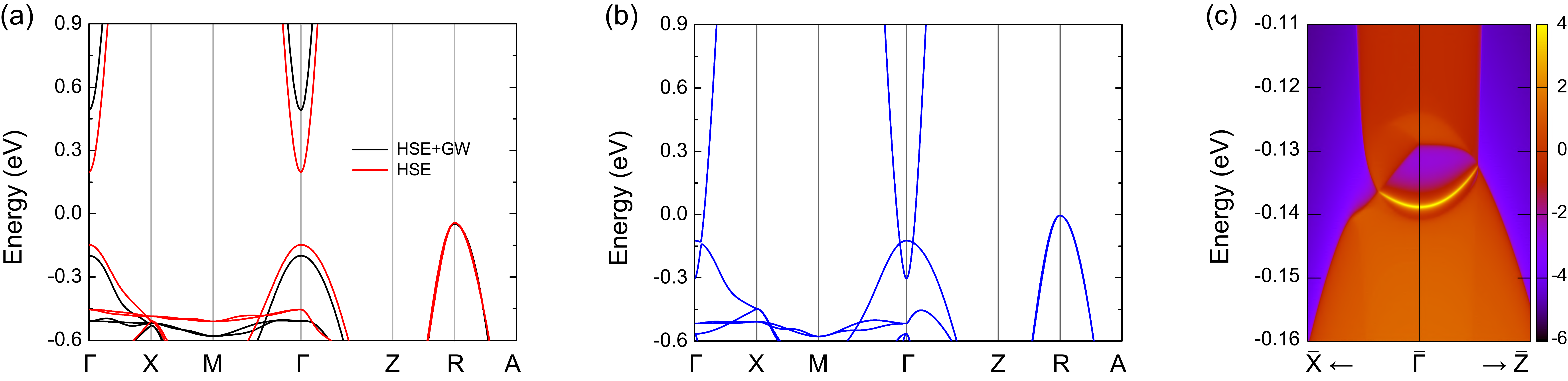}
\caption{(a) Electronic structures of unstrained $\beta$-PbO$_2$ using HSE06 and HSE06+$GW_0$. (b) Electronic structures of strained $\beta$-PbO$_2$. (c) The (010) surface states of strained PbO$_2$ using HSE06+$GW_0$. A warmer color represents a higher surface contribution.}
\label{f3-si}
\end{figure*}

As shown in Fig.\,\ref{f3-si}(a), the $GW_0$ band gap at the $\Gamma$ point increases to $0.69$ eV (the HSE06 band gap at the $\Gamma$ point is $0.34$ eV). Increasing the lattice parameters results in a decreasing band gap. As shown in Fig.\,\ref{f3-si}(b), under a modest tensile strain of $2.07$\% in the (\textbf{\textit{a}},\textbf{\textit{b}}) plane and $2.50$\% along the \textbf{\textit{c}} axis, the $GW_0$ band inversion energy reaches $0.18$\,eV, 
larger than the HSE06 band inversion of $0.11$\,eV under the same strain. These results demonstrate the robustness of the predicted temperature-promoted topological order in the presence of strong many-body interactions (in fact it becomes more prominent at the $GW_0$ level).

Using the HSE06+$GW_0$ band structures, we generate the Wannier functions arising from the $s$ orbitals of lead and the $p$ orbitals of oxygen. The $(010)$ surface states are calculated using an iterated surface Green's function method \cite{Wu2018}. As shown in Fig.\,\ref{f3-si}(c), the surface states are relatively dispersionless flat bands. These surface states with a divergent density of states can be detected by angle-resolved photoemission measurement \cite{Bian2016}, and may induce high temperature superconductivity \cite{Kopnin2011}.

\section{Finite temperature}

In this section we provide the details of the finite temperature analysis reported in the main text.

\subsection{Thermal expansion}

We calculate thermal expansion using the quasiharmonic approximation \cite{Dove1993}, in which we determine the equilibrium structure at temperature $T$ by minimizing the free energy at that temperature as a function of the $(\textbf{\textit{a}},\textbf{\textit{b}})$ and the $\textbf{\textit{c}}$ lattice parameters. We emphasize that all calculations need to be performed at the HSE06 level of theory to correctly reproduce the dynamical stability of $\beta$-PbO$_2$.

\begin{figure*}
\centering
\includegraphics[width=\linewidth]{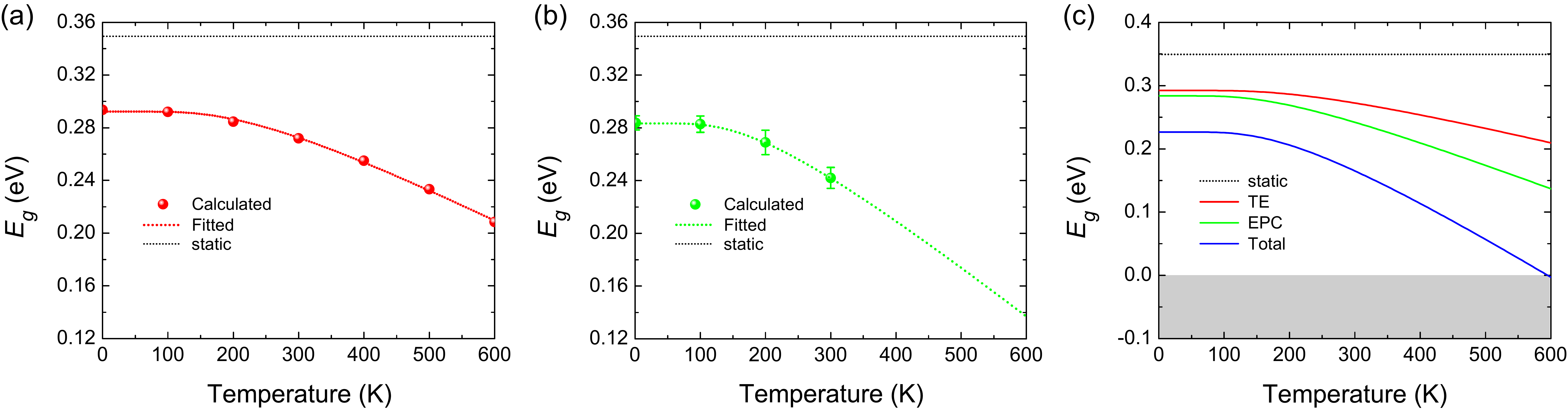}
\caption{Band gap of $\beta$-PbO$_2$ at the $\Gamma$ point as a function of temperature with contributions from (a) thermal expansion, (b) electron-phonon coupling, and (c) a combination of both.}
\label{f4-si}
\end{figure*}

Based on the temperature dependent volume of $\beta$-PbO$_2$, we calculate the change in the band gap arising from thermal expansion, as shown in Fig.\,\ref{f4-si}(a). Thermal expansion leads to a decrease in the band gap from $0.29$\,eV at $0$\,K to $0.21$\,eV at $600$\,K. We also note that the $0$\,K band gap of $0.29$\,eV is smaller than the static lattice band gap of $0.34$\,eV due to quantum zero-point motion, which also leads to lattice expansion. Taking the quantum zero-point motion into consideration, the calculated lattice constants at $0$\,K are \textbf{\textit{a}}$=4.988$\,\AA\ and \textbf{\textit{c}}$=3.373$\,\AA, while the static lattice constants are \textbf{\textit{a}}$=4.974$\,\AA\ and \textbf{\textit{c}}$=3.365$\,\AA.

We fit a Bose-Einstein oscillator model \cite{Monserrat2014} to the temperature-dependent band gap:
\begin{equation}
E_g(T) = A_0 + \frac{A_1}{ \textrm{e} ^{\overline{\omega}/k_{\mathrm{B}}T}-1},
\end{equation}
where $T$ is temperature, $k_{\mathrm{B}}$ is the Boltzmann constant, and $(A_0,A_1,\overline{\omega})$ are the fitting parameters. This model correctly reproduces the computational data, as shown in Fig.\,\ref{f4-si}(a). The $0$\,K value is given by $A_0=0.2923$\,eV, the coupling strength between band gap and volume is given by $A_1=-0.1802$\,eV, and the effective phonon frequency by $\overline{\omega}=59.80$\,meV.
\subsection{Electron-phonon coupling}

We calculate the change in the band gap of $\beta$-PbO$_2$ arising from electron-phonon coupling by evaluating the expectation value of the band gap with respect to the nuclear vibrational density. The expectation value is calculated using a stochastic sampling technique, accelerated by means of thermal lines \cite{Monserrat2016b,Monserrat2018}. The number of stochastic samples used is different at each temperature, and is chosen so as to reduce the statistical uncertainty to below about $5$\,meV.

All calculations are performed at the HSE06 level of theory using a $4\times4\times4$ supercell, and the total number of sampled configurations across all considered temperatures ($0$--$300$\,K) is over $300$.
We fit a Bose-Einstein oscillator model \cite{Monserrat2014} to the change of the band gap with temperature arising from electron-phonon coupling, as shown in Fig.\,\ref{f4-si}(b). The model allows us to extrapolate the calculated results up to $600$\,K, a necessary step because the computational expense of electron-phonon coupling calculations increases with increasing temperature due to the larger number of configurations that need to be considered at higher temperatures, making explicit calculations at the higher temperature range prohibitive.

As shown in Fig.\,\ref{f4-si}(b), including electron-phonon coupling leads to a $0$\,K band gap of $0.28$\,eV, which decreases to $0.17$\,eV at $600$\,K. For the Bose-Einstein oscillator fit, the $0$\,K band gap value is $A_0=0.2837$\,eV, the electron-phonon coupling strength has a fitted value of $A_1=-0.2234$, and the effective frequency is $\overline{\omega}=47.82$\,meV, corresponding to a phonon frequency of $385.45$\,cm$^{-1}$.

\subsection{Temperature dependence of the band gap}

The overall temperature dependence of the band gap of $\beta$-PbO$_2$ is a combination of thermal expansion and electron-phonon coupling. The combined results are shown in Fig.\,\ref{f4-si}(c), and show that the band gap at $0$\,K has a value of $0.23$\,eV, compared to a static lattice value of $0.34$\,eV. With increasing temperature, thermal expansion and electron-phonon coupling favor the topological phase. Taking both effects into consideration, a topological phase transition can be induced at $595$\,K. As discussed in the main text, the application of tensile strain enables the reduction of the transition temperature all the way to $0$\,K. 

\end{document}